\documentstyle[eqsecnum,floats,
preprint,
epsf,aps,tighten]{revtex}

\begin{document}

\preprint{\tighten \vbox{\hbox{CALT-68-2127} 
		\hbox{hep-ph/9707540} }}

\newcommand{\CP}{${\cal CP}$}
\newcommand{\C}{${\cal C}$}
\newcommand{\Parity}{${\cal P}$}
\newcommand{\B}{${\cal B}$}
\newcommand{\be}{\begin{equation}}
\newcommand{\ee}{\end{equation}}
\newcommand{\bea}{\begin{eqnarray}}
\newcommand{\eea}{\end{eqnarray}}
\title{Electroweak Baryogenesis Using Baryon Number Carrying Scalars}
\author{Hooman Davoudiasl\footnote[1]{hooman@theory.caltech.edu}, 
Krishna Rajagopal\footnote[2]{krishna@theory.caltech.edu~~~Address
after September 1, 1997: Center for Theoretical Physics, Massachusetts
Institute of Technology, Cambridge, MA 02139, USA.} and 
Eric Westphal\footnote[3]{westphal@theory.caltech.edu}}
\address{~\\Lauritsen Laboratory of High Energy Physics\\
California Institute of Technology, Pasadena, CA 91125, USA.}
\maketitle
\begin{abstract}
We describe a new mechanism for the generation of the baryon asymmetry
of the universe during a first order electroweak phase transition.  The
mechanism requires the existence of two (or more) baryon number
carrying scalar fields with masses and CP
violating mixing which vary with the 
Higgs field expectation value.
This mechanism can be implemented using squarks in
supersymmetric theories or using leptoquarks.
Our central observation is that 
reflection of these scalars from a bubble wall
can yield a significant net baryon number flux into the symmetric
phase, balanced by a flux of opposite sign into the
broken phase. 
For generic parameter choices,
scalars with incident energies in a specific, but not 
narrow, range yield order one
reflection asymmetries (between the probability of reflection of the
scalars and of their antiparticles).  The interesting 
energies are those for which there are two propagating scalars
in the symmetric phase but only one in the broken phase.
Electroweak sphaleron processes drive the baryon number
in the symmetric phase toward zero, but do not
act in the broken phase.  Our estimate of the
resulting baryon asymmetry is consistent with
cosmological observations for a range of mass 
parameters and CP violating phases in a supersymmetric implementation,
as long as the bubble walls are not too fast and not too thick.

\end{abstract}
\setcounter{page}{1}
\thispagestyle{empty}

\baselineskip 24pt plus 2pt minus 2pt

\section{Introduction}

Understanding how the baryon asymmetry of the universe (BAU) 
could have arisen from matter--antimatter 
symmetric initial conditions remains a central challenge
at the interface between high energy physics and cosmology.
The observed light element abundances, combined
with the theory of big bang nucleosynthesis, yield
the constraint that the ratio of the cosmological baryon
number density to the cosmological entropy density
lies in the range 
$3\times 10^{-11} < n_B/s < 9\times 10^{-11}$\cite{Copi}.
As has been understood since the work of Sakharov\cite{Sakharov},
a theory that explains the generation of
a net baryon asymmetry must include departures from
thermal equilibrium, and must
include processes that
violate baryon number conservation, charge conjugation symmetry C,
and CP symmetry, the product of charge conjugation and parity symmetry P.  
These conditions can be satisfied by electroweak processes
in the early universe, raising the possibility of an electroweak
explanation of the BAU[3-28].
In particular, Sakharov's requirements 
are all satisfied during a first order
electroweak phase transition[7-26].
As the universe cools through
the transition temperature $T_c$,
non-equilibrium conditions exist near
the bubble walls that are sweeping through the universe, converting the
high temperature ``symmetric'' phase to the low temperature ``broken'' 
phase.\footnote{By convention,
throughout the rest of this paper we refer to the phase
at $T=T_c$ that is smoothly connected to the equilibrium
phase at $T>T_c$
as the symmetric phase, and the phase at $T=T_c$ 
that is smoothly connected to the equilibrium phase at $T<T_c$
as the
broken phase, even though there is in fact no distinction between
the symmetry of the two phases.} 
Baryon number is violated in the electroweak theory
as a consequence of the anomaly\cite{tHooft} in processes in
which the sphaleron barrier is traversed\cite{sphaleron}.
The rate for baryon number violating processes is exponentially small
at zero temperature and in the broken phase
at the critical temperature 
$T_c$, but not in the symmetric phase\cite{KRS,AM}.
C is maximally violated in the electroweak theory, and 
CP violation enters via the Cabibbo-Kobayashi-Maskawa (CKM)
matrix.

Meeting Sakharov's conditions, as the minimal standard model
does, is necessary for generating a cosmological
baryon asymmetry, but need not
be sufficient for generating an asymmetry as large as that observed.
Indeed, in order to explain the observed BAU,
it appears that one needs to extend the standard model 
for at least two reasons.  First, although mechanisms
using CP violation from the CKM matrix alone have
been constructed\cite{F&S}, they seem to be unable to
generate the observed BAU\cite{H&S}. Hence, CP violation
beyond that in the standard model seems to be required.
Second, regardless of the mechanism
by which baryon number is generated, if it is to survive after
the electroweak phase transition, baryon number violating processes
must be sufficiently suppressed for $T<T_c$. This imposes a 
lower bound on 
the expectation value of the Higgs field in the broken
phase at $T=T_c$\cite{nowashout} which 
is not satisfied in the standard model with an experimentally
allowed Higgs mass.  
This bound can, however, be satisfied 
in various extensions of the standard model, including the minimal
supersymmetric standard model with
one top squark lighter than the top quark\cite{susynowashout}; 
throughout
this paper, we will assume that the theory
is suitably augmented such that the transition is 
strongly enough first order that the bound is satisfied.
Since the energies that are potentially relevant to electroweak 
baryogenesis will be increasingly explored in the coming decade by
present and planned accelerators, it is of great interest to explore
new weak-scale physics that can explain the observed BAU.

In this paper, we introduce a new mechanism for generating the
BAU at the  electroweak phase transition.  Our mechanism requires
augmenting the standard model by the addition of 
(at least) two baryon number carrying complex
scalar fields $\phi_1$ and $\phi_2$ with masses of ${\cal O}(T_c)$.
The two fields 
must be coupled by off-diagonal terms
in their (Hermitian) mass-squared matrix $M^2$ which include a CP violating 
phase. 
In this way, we introduce CP violation beyond that in the CKM
matrix.  Furthermore, we require that $M^2$ 
depend upon the Higgs field expectation values
in the theory so that the mass eigenvalues and eigenstates 
are different in the symmetric and broken phases.
The requirements just sketched can be implemented in
a variety of extensions of the 
standard model.  
Perhaps the most appealing  
possibility is that $\phi_1$ and $\phi_2$ are the 
$SU(2)$ singlet and $SU(2)$ doublet 
top squarks in a supersymmetric extension of the
standard model.  Another possibility, of interest
in light of the recent HERA anomaly\cite{HERA}, 
is that $\phi_1$ and 
$\phi_2$ may be weak-scale 
leptoquarks whose masses receive contributions from 
couplings to the Higgs field.  
Throughout this paper, we focus on the baryogenesis mechanism
rather than on model building.  
For definiteness, however, we present our mechanism and results 
taking the $\phi$'s to be squarks,\footnote{Previous treatments
of electroweak baryogenesis in supersymmetric theories
include those of Refs. \cite{CN,HN,susynowashout,macarena,aoki,worah}.} 
and defer discussion
of other possibilities to the concluding section.

In a supersymmetric theory,
the mass-squared matrix $M^2$ depends upon $v_1(T)$ and
$v_2(T)$, the temperature dependent expectation values 
of the two Higgs doublets $H_1$ and $H_2$ that give mass to
the down and up type quarks, respectively.  
The off-diagonal terms in $M^2$ are in fact zero in the symmetric
phase, where $v_1=v_2=0$.  In the broken phase, the off-diagonal
terms are complex and therefore CP violating. 
More generally, we define $\phi_1$ 
and $\phi_2$ as the eigenvectors of $M^2$ in the symmetric phase,
and note that the mass eigenstates in the wall and in the
broken phase are linear combinations thereof.

We will be interested in the reflection and transmission probabilities
of $\phi$'s incident on the wall from the symmetric phase. 
Because of the CP violating phases in $M^2$, the probability $R_{12}$
for an incident $\phi_1$ to be reflected back into
the symmetric phase as a $\phi_2$ is not the same as $R_{\overline{12}}$,
the probability for an incident antiparticle $\bar\phi_1$
to be reflected as a $\bar\phi_2$.  We will show that this reflection
asymmetry $\Delta R \equiv R_{12} - R_{\overline{12}}\,$ 
results in a net flux of baryon number from the bubble wall into the 
symmetric phase, compensated by a flux of the opposite sign into the
broken phase.  Note that the measure of CP violation
in the model is the spatial variation of the phase of the
off-diagonal terms in $M^2$.  A spatially constant phase can be rotated
away by a spacetime-independent unitary transformation on the
$\phi_i$; hence the phase must vary spatially 
if $\Delta R$ is to be nonzero.

The central observation of this paper is that the reflection asymmetry 
$\Delta R$ can be large, approaching 1, over a broad range of incident
energies, if the phase of 
the off-diagonal term in $M^2$ changes by ${\cal O}(1)$ as the bubble
wall is traversed from the symmetric phase into the broken phase.
As already noted, because
of the dependence of $M^2$ on $v_1$ and $v_2$,  the eigenvalues of
$M^2$ vary within the bubble wall.  We assume that the larger of the
two eigenvalues in the broken phase is greater than the larger of
the two eigenvalues in the symmetric phase.  This means that there is
in general a range of incident energies $E$ 
such that in the symmetric
phase, both eigenvalues of $M^2$ are below $E^2$, while in the broken
phase, there is only one eigenvalue below $E^2$.  Therefore,
there are two propagating modes with energy $E$ in the symmetric
phase and only one in the broken phase.  Consider a $\phi_1$ incident
upon the wall from the symmetric phase with an energy in this range.
As it begins to penetrate the wall, it evolves into a linear
combination of the position-dependent eigenstates of the matrix
$M^2$.  Since only one mode can propagate in the
broken phase, there is a position within the wall at which one mode is 
totally reflected.  The reflected mode, upon re-emerging into
the symmetric phase, is some linear combination of $\phi_1$
and $\phi_2$ which includes a significant $\phi_2$
component if there is significant mixing.
Another way of understanding what is special about
the range of energy under discussion is that at each
energy in this range, there is one linear combination
of incident $\phi_1$ and $\phi_2$ that is totally reflected.
For this reason, both $R_{12}$ and
$R_{\overline{12}}$ are generically large, and $\Delta R$ is also large
unless the CP violating phase is small.
We will refer to this range of incident energies as the
``enhanced reflection zone.''\footnote{We will show that for
energies in the enhanced reflection zone, scalars incident upon the wall
from the
the broken phase do not yield a reflection asymmetry and hence do not
contribute to the BAU.}
By comparison, for incident energies above the enhanced
reflection zone, for which
there are two propagating modes in both the symmetric and broken phases
and throughout the bubble wall, we find that $\Delta R$ is nonzero
but is generically many orders of magnitude smaller than one.
The width in energy of the enhanced reflection zone is comparable to
the amount by which the masses change between the two phases; in the
example we present, the width of the enhanced reflection zone is $90$ GeV.
In Section II, we present the parametrization of $M^2$ appearing in a
supersymmetric theory.  We then set up the calculation of $\Delta R$,
leaving a detailed presentation of the method of calculation to the
Appendix.  We evaluate $\Delta R$ and explore its dependence on
parameters in $M^2$.  

Standard model quarks also have an enhanced reflection zone, as
discussed by Farrar and Shaposhnikov 
\cite{F&S}, although its width is only of order the strange quark
mass.  The resulting BAU is small \cite{H&S}, essentially because the light
quarks have mean free paths much shorter than their Compton
wavelengths.  We defer to Section IV
a discussion of the suppression due to the finite mean free path of
the heavy scalars we employ in our mechanism; the suppression is not severe.

In Section III, we integrate $\Delta R$ against the appropriate
thermal distributions for incident $\phi_1$'s and $\phi_2$'s to
obtain the baryon number flux injected into
the symmetric phase.   If the wall velocity $v_w$ is zero,
or if the masses of $\phi_1$ and $\phi_2$ in the symmetric phase
are equal, we find that the baryon number flux due to incident
$\phi_1$'s is cancelled by that due to incident $\phi_2$'s.  As long 
as $v_w\neq 0$ and the masses in the symmetric phase are
not degenerate, we obtain a nonzero baryon number flux.
The larger the fraction of the thermal
distributions for incident $\phi$'s lying in
the enhanced reflection zone, the larger the baryon number
flux will be.  The final element in the mechanism
involves electroweak baryon number violating processes.  These
drive the baryon number density in the symmetric phase
toward zero.  Because they do not act in the broken
phase, the final result is a net baryon asymmetry of the universe
whose magnitude we estimate in Section IV.  
Our mechanism yields a BAU consistent with observation
if the scalars have nondegenerate masses of order $T_c$ in the
symmetric phase and if the bubble walls are sufficiently thin and slow.
We discuss open
questions and model implementations in Section V, and note there
that an enhanced reflection zone can arise in
leptoquark models, and thus is not peculiar
to supersymmetric theories.

To close this introduction, we contrast our mechanism 
with the charge transport mechanism, pioneered by Cohen,
Kaplan and Nelson\cite{CKN2,CKN4},
further developed by many authors, and 
used to estimate the BAU generated 
during the electroweak phase transition in supersymmetric 
theories[19,23-26].
Our mechanism can be seen as a modification of the charge transport
mechanism.  We make explicit comparisons with 
the results of Huet and Nelson\cite{HN} obtained using
the charge transport mechanism and
find that our mechanism can yield an
$n_B/s$ consistent with observation for smaller 
CP violating phases.
The central difference is that in our mechanism, we generate
a flux of baryon number into the symmetric phase,
whereas in the charge transport mechanism a flux
of another quantum number, often left-handed
baryon number minus right-handed baryon number,
is generated.  This axial baryon number can be washed out by QCD processes
before it has time to bias electroweak baryon
number violating processes\cite{GS}.    
Because our mechanism generates a baryon
number flux, it is immune to QCD interference of this
kind.\footnote{We should note that there are scenarios in which baryogenesis
via the generation of an axial baryon number current can be immunized
against suppression due to strong sphalerons.  One example\cite{HN}
requires that the left- and right-handed top squarks and the left-handed
bottom squark have symmetric phase masses comparable to the temperature
while the other squarks are heavier.  Another example\cite{moore}
requires the formation of a squark condensate just above $T_c$.}
This contrast is particularly germane in light of
the recent demonstration that the rates for the
relevant QCD processes are significantly larger than
previously expected\cite{moore}.  Various authors have noted the
possibility that a baryon number flux may be generated,
but this has always been assumed to be a small effect.  This is in fact
true for incident energies such that the number of propagating
modes is the same on both sides of the bubble wall.   For a generic 
mass-squared matrix $M^2$, however, there is a broad region of 
incident energies in which fewer modes propagate in the
broken phase. We observe that this leads to large reflection asymmetries,
and consequently to a large baryon number flux into the 
symmetric phase, yielding an efficient mechanism for
generating a BAU consistent with cosmological observations
during the electroweak phase transition.
We will refer to the electroweak baryogenesis mechanism we
propose as 
the scalar baryon number transport mechanism.

\section{$M^2$, $\Delta R$, and the Enhanced Reflection Zone}

We begin this section by presenting
the parametrization of $M^2$ appropriate when $\phi_1$ and $\phi_2$
are left- and right-handed top squarks and 
set the stage for the
calculation of $\Delta R$.  We then describe the dependence of
$\Delta R$ upon the incident energy and upon parameters in $M^2$.

As discussed in the introduction, the CP violation that we 
exploit results from the mixing between two scalars
as they traverse the bubble wall  separating regions 
of symmetric and broken phase at $T=T_c$.
In this background, terms in the potential
that couple the baryon number carrying scalars $\phi_i$ to the Higgs
fields $H_1$ and $H_2$ give the $\phi_i$
spacetime--dependent masses and mixings which can be encoded
in a $2 \times 2$ Hermitian mass matrix.  
In a supersymmetric theory, the 
scalars $\phi_i$ are squarks.
Top squarks (stops) are the most promising candidates
to play the role we envision for the $\phi_i$'s 
because they can be light
without violating experimental upper bounds 
on neutron and electron electric dipole moments (EDMs).  Indeed,
Cohen, Kaplan, and Nelson have recently advocated supersymmetric
models in which CP violating phases are ${\cal O}(1)$, but 
observable EDMs do not arise because the first and second
generation squarks have masses in the tens of TeV\cite{nonMSSM}.
Although it is conceivable that the scalar baryon number transport mechanism
could be implemented using first or second generation squarks,
it seems likely that stops will yield the largest
contribution to the BAU.
The stop mass matrix is\cite{mark}
\begin{eqnarray}
M^2=\left( \begin{array}{cc}
~~\tilde{m}_{tR}^2 +m_t^2 + m_Z^2 \cos 2 \beta 
\left( \frac{2}{3}\sin^2 \theta_W \right)~~ & 
m_t (A e^{ i \varphi_A} + \mu e^{ i \varphi_B} \cot \beta) \\~&~\\
m_t (A e^{-i \varphi_A} + \mu e^{-i \varphi_B} \cot \beta) &
~~\tilde{m}_{tL}^2+m_t^2  
+ m_Z^2  \cos 2 \beta \left(\frac{1}{2}-\frac{2}{3} \sin^2 \theta_W\right)~~
\end{array} \right). \label{squarkmass}
\end{eqnarray}
Here, $m_t=\lambda_t v_2$ is the top quark mass, 
$m_Z^2=g^2(v_1^2 + v_2^2)/2$ is the $Z$-boson mass squared,
$\tilde{m}_{tR}$
is the soft SUSY-breaking mass for the 
$SU(2)$ singlet stop, $\tilde{m}_{tL}$ is the
soft SUSY-breaking mass for the $SU(2)$ doublet
stop, $\tan \beta=v_2/v_1$ is the ratio
of the Higgs field expectation values, $\mu e^{i \varphi_B}$ is the
(complex) mass in the Higgs potential coupling the two Higgs fields,
and $A e^{i \varphi_A}$ is a complex soft SUSY-breaking
term.  We will see that the scalar baryon number transport mechanism
works best if
$\tilde{m}_{tL}$
and $\tilde{m}_{tR}$ are both ${\cal O}(T_c)$ and differ by
about 10-30\%.   Mass differences
of this order
can arise due to renormalization group evolution down from
some high energy scale at which $\tilde{m}_{tL} =\tilde{m}_{tR}$. 
Indeed, in the models of Ref. \cite{falk}, $\tilde{m}_{tL}$
and $\tilde{m}_{tR}$ differ by 20\%. 

During most of the existence of the expanding bubble, its wall can be
treated as flat because its radius of curvature is much larger than its
thickness, so the mass matrix will depend only
upon one spatial direction, which we take to be the $x$ direction.  We choose
the convention that the region of large negative $x$ is the symmetric
phase and that of large positive $x$ is the broken phase.  
Both $v_1$ and $v_2$ vary across the bubble wall.  
A complete
calculation of these profiles is beyond the scope of this paper,
although a treatment using the resummed one-loop temperature-dependent
effective potential is possible\cite{vilja}. 
As is conventional, we make a simple choice in terms of a
single width
parameter, hoping that this captures the essential physics.
Following Ref. \cite{macarena}, we choose profiles such 
that the $v_i(x)$ are $x$-independent for $x<-w/2$ and $x>w/2$
and are sinusoids for $-w/2<x<w/2$.  
That is, we define the profile function
\begin{equation}
p(x)=\left\{ \begin{array}{ll} 0 & x\leq -w/2\\
\frac{1}{2} + \frac{1}{2}\sin\left(\frac{\pi\,x}{w}\right) & -w/2 < x < w/2\\
1 & x\geq w/2 \end{array}\right.
\label{profile}
\end{equation}
and then for 
$v_1$, we take
\begin{equation}
v_1(x) = v_1^0 \, p(x)\ ,
\label{v1profile}
\end{equation}
which varies smoothly from zero in the symmetric phase
at $x\rightarrow -\infty$ to $v_1^0$ 
in the broken phase at $x\rightarrow \infty$.  The parameter
$w$ characterizes the width of the wall separating the two
phases.  
Although in reality the profiles
behave exponentially for $|x|\gg w$, most of the interesting
physics happens where $x$ is varying most rapidly and
$p(x)$ is as good a parametrization as any. 
The choice (\ref{profile}) is convenient
numerically, as it allows us to impose boundary conditions
at $x=\pm w/2$ rather than at larger $|x|$.  

It is crucial to our mechanism
that $v_2/v_1$ vary across the bubble wall so that the overall phase
of the off-diagonal term in $M^2$ is not constant.
We take 
\begin{eqnarray}
v_2(x) &=& v_1(x) \tan\beta(x)\nonumber\\
&=&v_1(x) \left[ p(x)\,\tan\bar\beta
+ \left(1-p(x)\right)\,\tan(\bar\beta - \Delta\beta)\right]
\label{betaprofile}
\end{eqnarray}
so that $\tan\beta(x)$ varies from $\tan(\bar\beta-\Delta\beta)$ in the 
symmetric phase to $\tan\bar\beta$ in the broken phase.
We will take $\tan\bar\beta = 2$ in our estimates. Our results
do not depend sensitively on this choice.
The appropriate choice for $v_1^0$ in (\ref{v1profile}) is {\it not} 
$(250~{\rm GeV})\cos \bar\beta$, the value it takes at $T=0$, but 
rather the value it takes
in the broken phase at $T=T_c$.
This can be calculated as a function of parameters
in specific models, but we will
simply use the reasonable estimate 
$v_1^0 = (2/3)(250~{\rm GeV})\cos\bar\beta$.
(Our results do not depend sensitively on the choice of
prefactor.) In order for the baryon
asymmetry generated (by any mechanism)
during the electroweak phase transition
not to be wiped out, $(v_1^0)^2+(v_2^0)^2$ must be larger than $T_c^2$.
Estimates for $\Delta \beta$ exist in specific models 
and range from $0.01-0.03$\cite{macarena,quiros}
to $0.25$\cite{worah}, but there are 
certainly no experimental constraints on this parameter.
In order for the overall phase of the off-diagonal terms in
$M^2$ to vary with $x$, that is, in order for $M^2$ to
introduce CP violating effects,
we must have 
\begin{eqnarray}
\Delta \beta&\neq& 0\nonumber\\
\Delta \varphi &\equiv& \varphi_A - \varphi_B\neq 0\ .
\label{deltavarphi}
\end{eqnarray}  
EDM experiments may constrain $\Delta\varphi$
in some models\cite{susyedm}, but it has recently been noted\cite{falk}
that in other models $\varphi_B$ is constrained to
be small while $\varphi_A$ is essentially unconstrained.
Any
constraints on the CP violating phases are weakened
if the first 
and second generation squarks are heavy. Finally,
note that $\varphi_A$ can be generation-dependent.
Hence, there is no model-independent constraint on $\Delta\varphi$
for third generation squarks.

\begin{figure}
\centerline{
\epsfysize=3.5in
\epsfbox{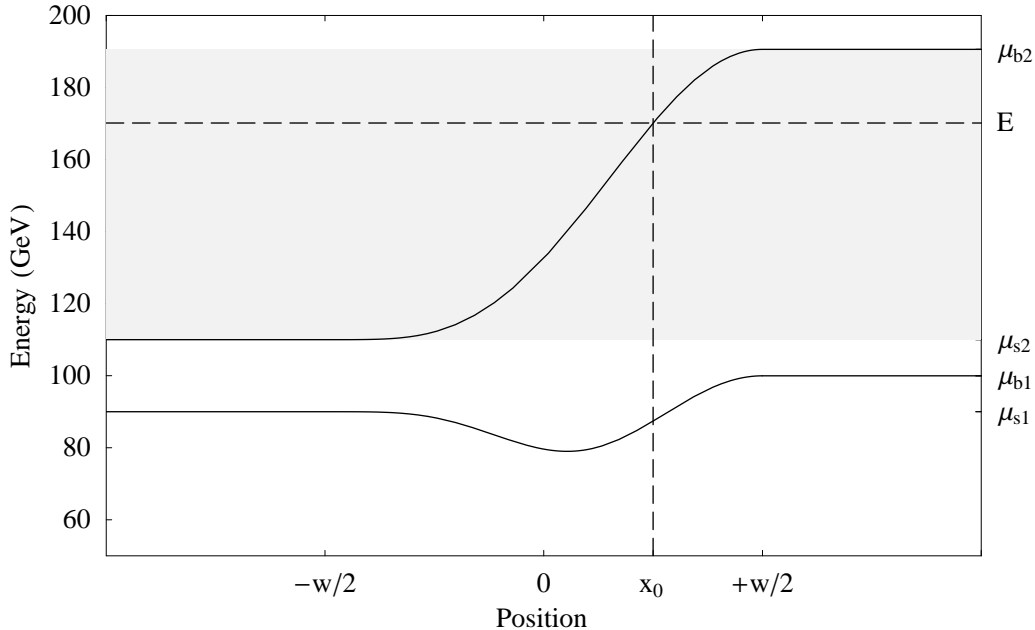}}
\caption{Mass eigenvalues as a function of position. The two curves
show the square roots of the eigenvalues of $M^2$, with parameters
given in the text, as a function of $x$.  $w$ is the wall width.
$\mu_{s1}$ and $\mu_{s2}$ are the mass eigenvalues in the symmetric phase,
and 
$\mu_{b1}$ and $\mu_{b2}$ are the mass eigenvalues in the broken phase.
For any incident energy in the shaded 
range $\mu_{s2}<E<\mu_{b2}$, there
is a position $x_0$ at which the upper eigenvalue crosses $E$.
}
\label{fig1}
\end{figure}
As a concrete example which will serve as a visual aid for 
much of our subsequent discussion, in Figure 1,
we
plot the eigenvalues of $M^2$ as a function of $x$
for the parameter set $\tilde m_{tL} = 110 ~{\rm GeV}$,
$\tilde m_{tR} = 90 ~{\rm GeV}$, $A=\mu=100 ~{\rm GeV}$,
$\tan\bar\beta=2$, $\tan(\bar\beta-\Delta\beta) = 1$, that
is $\Delta \beta=0.32$, 
$\Delta\varphi = \pi/2$,
$m_Z=91 ~{\rm GeV}$, $m_t=175~{\rm GeV}$, and
$\sin^2 \theta_W = 0.23$.
With these parameters, the zero temperature masses
of the two top squarks are $141~{\rm GeV}$ and $243~{\rm GeV}$.  
In our example, we take the wall width to be $w=(4~{\rm GeV})^{-1}$.
Although the critical temperature $T_c$ 
and the wall velocity $v_w$ play no role in the calculation
of $\Delta R$, we mention for completeness that when they
enter in the next section, we will use $T_c=100~{\rm GeV}$
and $v_w=0.1$ as representative values. 
It is conventional to write the wall width in terms of the
temperature, that is, $w=25/T_c$ in our example.  Henceforth,
we write $T_c$ as $T$ when it causes no ambiguity.  
The final relevant parameter is the $\phi$
mean free path $l$, which will appear in Section IV
where we estimate $l\sim 10/T\sim 0.4 w$.
This completes the enumeration of the parameters specifying
the ``canonical'' example for which we will quote results
in the following.  We will, of course, describe the effects
of varying each of these parameters at appropriate points
in the discussion.

We wish to follow a $\phi$
particle from the thermal ensemble in the symmetric phase that last
scattered somewhere away from the wall and is now incident upon
the wall.
Implicit in this
scenario is the assumption that the mean free path of the scalars in
the plasma is long compared to the wall width $w$.
This assumption is likely false, but we 
nevertheless use it in this section and the next, deferring 
our treatment of the suppression due to finite mean free paths for
the $\phi$ particles 
to Section IV.
The particle impinges
upon the wall and is reflected or transmitted, and then resumes
its thermal motion in
the plasma on one side of the wall or the other.  
During the time between the last scattering before reflection
or transmission and the first after, the particle propagates
freely, feeling only the changing expectation values of the Higgs fields 
which are
encoded in the mass matrix.  
We calculate the reflection coefficients and their asymmetry 
in the rest frame of the wall;
we will boost the resulting baryon number flux to the 
plasma frame when we calculate it in Section III.
In the wall frame, energy is conserved upon traversing the wall since
the mass matrix is time-independent.  
The reflection coefficients can be calculated by
solving
the time independent Klein-Gordon equations
\begin{equation}
\left[ \delta_{ij} (\nabla^2 + E^2) - M^2_{ij}({\bf x}) \right] 
\phi_j({\bf x}) = 0 
\label{kg}
\end{equation}
because the time dependence of solutions is simply an overall 
$\exp(iEt)$.
In general, we will take a
basis in $\phi_1$ and $\phi_2$ such that far from the wall in the 
symmetric phase, $M^2$ is diagonal.  This has
already been accomplished, since $\tilde m_{tL}^2$ and $\tilde m_{tR}^2$
are the only terms in (\ref{squarkmass})
which are nonzero in the symmetric phase.
When calculating the baryon number flux in Section III,
we will consider particles incident upon the wall
with momenta that are not perpendicular to the wall. However, their
reflection coefficients will depend only upon the component of their
momentum that is perpendicular to the wall; hence, it suffices 
to compute reflection coefficients for normal incidence. 
Therefore, the problem of finding reflection coefficients is effectively
a one-dimensional scattering problem, and equations (\ref{kg}) become
ordinary differential equations for $\phi_j(x)$.

Consider the mass matrix whose eigenvalues are shown in Figure 1.
The behavior of the reflection coefficients is qualitatively
different for particles incident from the
symmetric phase with 
energies $E<\mu_{s2}$,
$\mu_{s2}<E<\mu_{b2}$, and $E>\mu_{b2}$.
For $E<\mu_{s2}$, there is only one propagating mode in
both the symmetric and the broken phases.  
In general, we denote the
reflection coefficient for a $\phi_i$ from the
symmetric phase reflected back into the symmetric phase as a
$\phi_j$ by $R_{ij}$ for particles,  and
denote the corresponding quantity for antiparticles by
$R_{\overline{ij}}$. 
For $E<\mu_{s2}$, however, the only reflection coefficients are 
$R_{11}$ and $R_{\overline{11}}$.
Since the CPT conjugate of the reflection
of $\phi_1$ to $\phi_1$ is the reflection of
$\bar{\phi}_1$ to $\bar{\phi}_1$, 
we see that
$R_{11}=R_{\overline{11}}$, and there is no reflection
asymmetry.  Before proceeding to higher energies, note 
that for a different mass matrix it may be the case
that $\mu_{b1}>\mu_{s2}$.  In this case, for 
$E<\mu_{b1}$ there are two propagating modes 
in the symmetric phase and none in the broken phase,  
so both $\phi_1$ and $\phi_2$ must
be totally reflected.  We now argue that in this circumstance,
the reflection coefficients again cannot be CP violating. 
Unitarity implies that for total reflection,
\begin{eqnarray}
R_{11}            + R_{12}            &=& 1 \label{unitaritypos} \\
R_{\overline{11}} + R_{\overline{12}} &=& 1 \label{unitarityneg}
\end{eqnarray}
We see that unitarity, together with $R_{11}=R_{\overline{11}}$, 
implies that $R_{12}=R_{\overline{12}}$ and hence there is no reflection
asymmetry. 
For the rest of this paper, it is implicit that references
to $\mu_{s2}$ should be replaced by references to $\mu_{b1}$
if the mass matrix is such that $\mu_{b1}>\mu_{s2}$.
We have shown that 
particles incident from the symmetric phase 
with energy
$E<\mu_{s2}$
yield no reflection asymmetry.

Now consider incident energies $\mu_{s2}<E<\mu_{b2}$, for
which there are two propagating modes in the symmetric phase
and only one in the broken phase.
As shown in Figure 1, for any energy in this shaded range
there is a point $x_0$ at which the larger eigenvalue
of $M^2$ equals $E^2$. This means that there is a particular
linear combination of incident $\phi_1$ and $\phi_2$
that evolves by mixing as it propagates through the wall in just such
a way that upon arrival at $x_0$ it is purely in
the mass eigenstate with eigenvalue $E^2$, and  
is therefore totally reflected.  
Since
one linear combination
of $\phi_1$ and $\phi_2$ is totally reflected
by the wall, 
both $R_{12}$ and
$R_{21}$ are large in this energy range, given
sufficient mixing.
In order to obtain large asymmetries between the reflection
coefficients for particles and for antiparticles, 
the individual reflection coefficients must of course be large,
making this zone of enhanced reflection a promising place
to look for large $\Delta R$.
Without CP violation, the same linear combination of incident
$\bar\phi_1$ and incident $\bar\phi_2$ mixes
to become the mode with eigenvalue $E^2$ at 
$x_0$, and is totally reflected.  This implies that
$R_{\overline{12}}=R_{12}$ and
$R_{\overline{21}}=R_{21}$.  However, 
if the
off-diagonal term in $M^2$ has a spatially varying
phase, then the linear combination of $\bar\phi$'s
that is totally reflected is different
from that for the $\phi$'s, and we expect $\Delta R\neq 0$,
as we confirm explicitly below.

Before proceeding, 
note that
$R_{\overline{12}}=R_{21}$ 
by CPT and therefore 
\begin{equation}
\Delta R \equiv R_{12} - R_{\overline{12}}= R_{12}-R_{21}\ .
\end{equation}
This simplifies calculations by giving the reflection asymmetry
in terms of reflection coefficients for particles only.
Another simplification is that in the enhanced reflection zone,
$\Delta R$ is the only possible asymmetry because
there is no asymmetry
due to particles incident from the broken phase. Denoting
the single propagating mode in the broken phase by $\phi_3$,
since $R_{33}=R_{\overline{33}}$ by CPT, unitarity requires
that incident $\phi_3$'s cannot yield baryon number
asymmetric transmission into the symmetric phase.
We give a careful explanation of the calculation
of $\Delta R$ in the Appendix; henceforth in this section,
we focus solely on the results.

\begin{figure}
\centerline{
\epsfysize=3.5in
\epsfbox{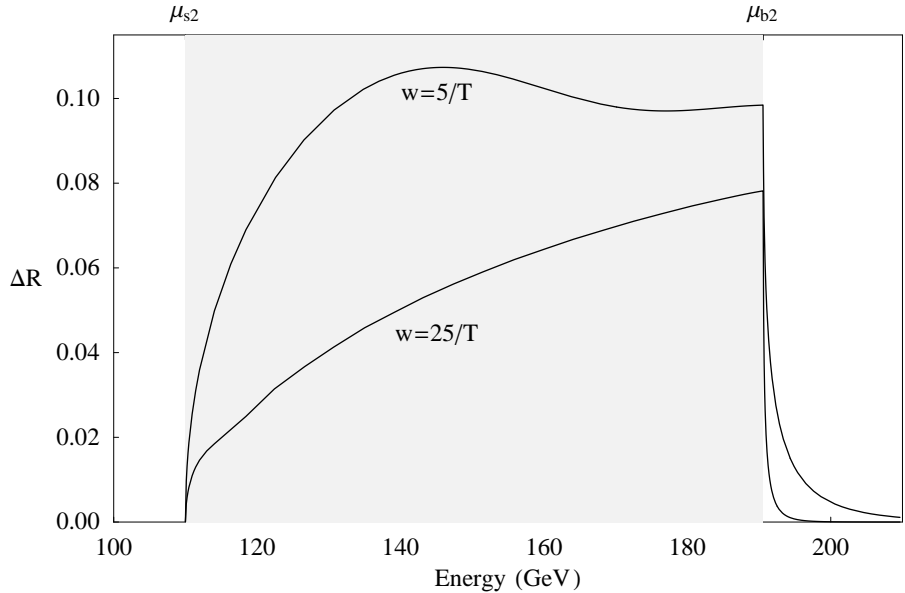}}
\caption{$\Delta R$ as a function of incident energy $E$, for two
different wall widths $w$.  The enhanced reflection zone is
shaded.}
\label{fig2}
\end{figure}
In Figure 2, we plot $\Delta R$ vs. incident energy $E$
for two values of the wall width, $w=5/T$ and $w=25/T$,
with all other parameters as in our canonical 
example.  The enhanced reflection zone is apparent, with
$\Delta R \sim 0.1$ for $w=5/T$.  Even greater values of
$\Delta R$ are obtained
for $\Delta \beta$ and $\Delta\varphi$ that are larger than
our canonical values.  (For example, for $\Delta\beta=1$
and $\Delta\varphi=2.5$, we find $\Delta R\sim 0.3$.)
We have therefore confirmed
that in the enhanced reflection zone, large reflection
coefficients yield large reflection asymmetries in the
presence of ${\cal O}(1)$ CP violating phases.
For incident energies below the enhanced reflection
zone, $\Delta R\equiv 0$.  At higher energies, that is for 
$E>\mu_{b2}$, there
are two propagating modes in both the symmetric and the broken
phases and throughout the wall.
In this regime,  $\Delta R\neq 0$, but it
is extremely small.  For $w=5/T$,  $\Delta R=0.0047$ at $E=200~{\rm GeV}$;
$\Delta R=2.5\times 10^{-5}$ at $E=250~{\rm GeV}$; and
$\Delta R=1.3\times 10^{-6}$ at $E=300~{\rm GeV}$.  For these energies, 
transmission coefficients are very close to $1$ and
reflection coefficients (and therefore $\Delta R$) 
are very small.  

For energies above $\mu_{b2}$, unlike for those in the 
enhanced reflection zone, there
is the possibility of a nonzero asymmetry arising
from particles and antiparticles incident from 
the broken phase and transmitted through the wall into the symmetric
phase. This asymmetry is very small, as we now argue.
Just as the $R_{ij}$ are very small in this energy range,
the reflection coefficients for reflection of particles
incident from the broken phase back
into the broken phase are very small also.  Since any 
asymmetry associated with transmission into the
symmetric phase must be balanced by an asymmetry
in reflection, we conclude that even though
the transmission coefficients are $\sim 1$, their
asymmetry is very small.  

\begin{figure}
\centerline{
\epsfysize=3.5in
\epsfbox{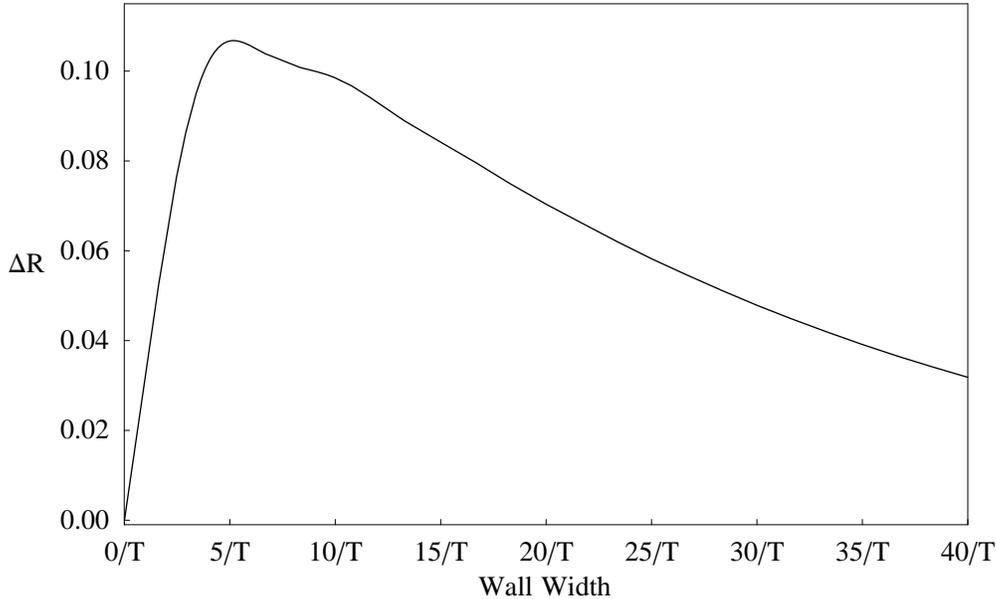}}
\caption{$\Delta R$ as a function of wall width $w$
for incident energy $E=150~{\rm GeV}$, which is within the 
enhanced reflection zone.}
\label{fig3}
\end{figure}
We now discuss the dependence of $\Delta R$ on the
parameters in the problem, beginning with the wall width
$w$.
Note that in order to obtain a nonzero $\Delta R$,
there must be some region in $x$
in which modes incident from the
symmetric phase can mix, and hence feel the 
effects of the CP violating terms, 
before arriving at $x_0$ where one
mode is totally reflected.
This implies that $\Delta R=0$ for an infinitesimally thin (step function)
wall. 
In the thin wall limit, $R_{12}$ and $R_{\overline{12}}$ are large
in the enhanced reflection zone, but they are equal.
In Figure 3,
we plot $\Delta R$ at $E=150~{\rm GeV}$  versus the wall width $w$
for our canonical $M^2$.   
We see that
$\Delta R \sim w$ for $w$ small relative to the
inverse mass scales in the problem.  At large $w$, the number
of wavelengths per wall width grows and therefore 
$R_{12}$ and $R_{\overline{12}}$ decrease,
and so does
$\Delta R$. 
The asymmetry $\Delta R$ peaks at about
$w=5/T$, one of the values we have
chosen to plot in Figure 2.  This is an unphysically thin wall ---
estimates for $w$ range from $10/T$ to $100/T$.
(The physics determining $w$ is presented, for example, in Refs.
\cite{linde,mp}.)
In
our canonical example we follow Ref. \cite{macarena} and
use $w=25/T$, and we have plotted $\Delta R$ for this 
wall width in Figure 2.  If $w$ is in fact smaller
than $25/T$, our final result for the BAU is enhanced, while for 
thicker walls, it is somewhat suppressed.

Let us now consider the effects of varying the mass parameters in 
$M^2$.
We define $m$ and $\Delta m$ through
\begin{eqnarray}
\mu_{s1} = \tilde m_{tR} &=& m - \frac{\Delta m}{2}\nonumber\\
\mu_{s2} = \tilde m_{tL} &=& m + \frac{\Delta m}{2}\ .
\end{eqnarray}
For simplicity, we will always take $\mu=A$ and $\tan\bar\beta = 2$.  
(The optimal choice for $\bar\beta$ should be such that
$A\sim\mu\cot\bar\beta$. We find, however, that for
$A=\mu$, varying $\bar\beta$ from 0.5 to 4 changes our results
by at most 10\%, so the dependence on $\bar\beta$ is not significant.)
We have investigated the dependence of our results on
$m$, $\Delta m$ and $A$.
Of course, we
do not vary the $m_Z^2$ and $m_t^2$ terms in $M^2$, and
these should be thought of as setting the energy scale.
Varying $m$ between 
$50$ and $150~{\rm GeV}$ while holding $\Delta m/m$ and $A$
fixed changes $\Delta R$ 
by less than 20\%.  Increasing $m$ further leads to a suppression
of $\Delta R$
because it increases all the eigenvalues relative to $1/w$.
For $A=0$, there is no mixing between $\phi_1$ and $\phi_2$,
and $\Delta R=0$.   Increasing $A$ from $0$ to $200~{\rm GeV}$
holding all else fixed yields
a monotonically increasing $\Delta R$, but in going
from $A=100~{\rm GeV}$ to $A=200~{\rm GeV}$, 
the increase in $\Delta R$ is less
than 10\%.  Of the three mass parameters we have varied, $\Delta m$
has the biggest effect.  $\Delta R$ is maximized for $\Delta m =0$.
However, we will see in the next section that the baryon number flux
is identically zero for $\Delta m=0$.  As $\Delta m$ is increased,
$\Delta R$ falls; by a few percent at $\Delta m/m=0.2$; by about a
factor of two at $\Delta m/m = 0.8$.  
We defer a plot displaying the dependence of our results on
$\Delta m$ to the next section.

Finally, we come to the CP violating phase.  We have verified that
for small $\Delta\varphi$ and $\Delta\beta$, $\Delta R$ is linear
in both quantities.
As we have noted, there is no model-independent constraint on $\Delta
\varphi$; therefore, in Figure 4 we show $\Delta R$ over the entire
range of $\Delta \varphi$.
We see that $\Delta R$ is approximately linear in $\Delta \varphi$ for
$| \Delta \varphi | \leq 2$.
We find that $\Delta R$ is linear
in $\Delta \beta$ over the range $-0.2<\Delta \beta<0.5$.
The values we have been using in our canonical example ---
$\Delta\varphi=\pi/2$ and $\Delta\beta=0.32$ --- are within
the linear regime.  For convenience, in subsequent sections, we
often quote results assuming that $\Delta \beta$ and $\Delta \varphi$
are in their linear regimes, as is likely the case for $\Delta \beta$,
but not necessarily for $\Delta \varphi$.
Fortified by our understanding of how $\Delta R$ in the enhanced
reflection zone varies with wall width, masses, and phases, we
are ready for the next step in our computation of the BAU.
\begin{figure}
\centerline{
\epsfysize=3.5in
\epsfbox{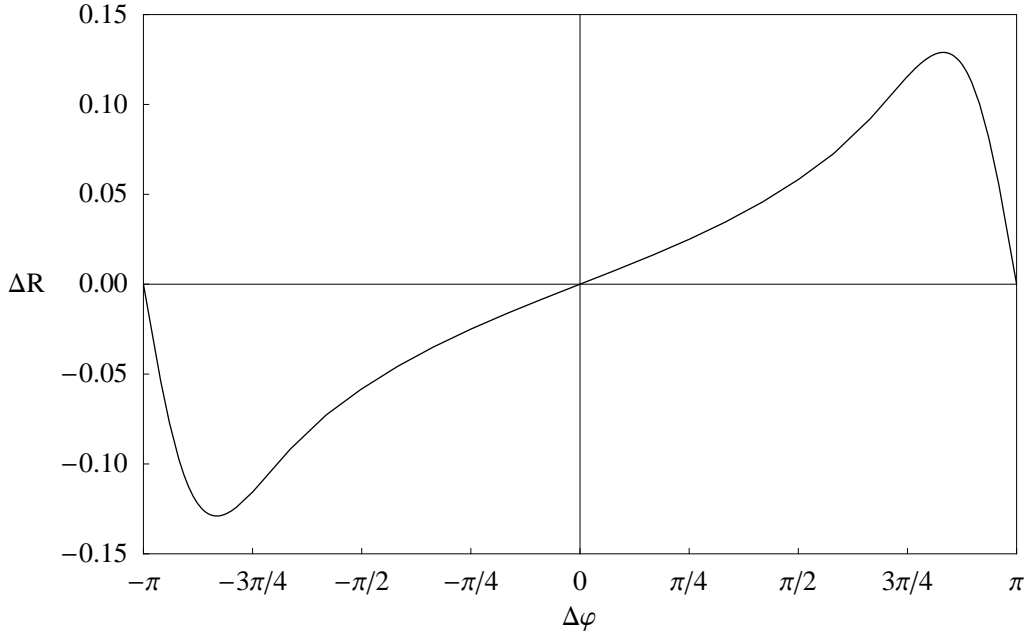}}
\caption{$\Delta R$ as a function of $\Delta \varphi$ 
for incident energy at the center of the
enhanced reflection zone.}
\label{fig4}
\end{figure}

\section{From Asymmetric Reflection Coefficients to  Baryon
Number Flux}

We now compute the baryon number 
flux into the symmetric phase by integrating 
$\Delta R$ against the flux densities of the incident particles.  
The flux, which we denote by ${\cal F}_B$, receives two contributions.  
One contribution, $F_{s \to s}$, is generated by the asymmetry 
in the reflection of particles and antiparticles 
incident on the wall from the symmetric phase
and is associated with $\Delta R$.  
The other contribution, $F_{b \to s}$, is caused by the asymmetry in 
the transmission of the particles and their antiparticles incident 
on the wall from the broken phase.  We have seen, however, 
that this asymmetry (unlike $\Delta R$) is zero for energies in the enhanced
reflection zone, and is small (like $\Delta R$) at higher
energies.  We therefore neglect $F_{b \to s}$ and
obtain 
\begin{equation}
{\cal F}_B \simeq \frac{1}{3}F_{s \to s} \ .
\label{FB}
\end{equation}
The factor of 1/3 arises because the $\phi$ particles have baryon number 1/3.

Before deriving an expression for $F_{s \to s}$, we 
make some general observations.
The processes we are studying are invariant under time translations
(although not under time reversal!).  
Therefore, energy is conserved upon reflection from 
or transmission through the wall.
We denote the components of a particle's momentum in the 
rest frame of the wall that are perpendicular and parallel
to the wall by
$p_{i\perp}$ and ${\vec p_{i\parallel}}$, respectively.
Recalling that $M^2$ is diagonal in the symmetric phase, 
we have 
\begin{equation}
E^2 = p_{1\perp}^2 + p_{1\parallel}^2 + \mu_{s1}^2 = 
p_{2\perp}^2 + p_{2\parallel}^2 + \mu_{s2}^2\ ,
\label{E}
\end{equation}
where $p_{i\parallel} = |{\vec p_{i\parallel}}|$.
Since the processes of reflection from and transmission through
the wall are invariant under spatial translations in
the directions parallel to the wall, $\vec p_{i\parallel}$ 
is conserved.  For example, for a $\phi_1$ reflecting into a $\phi_2$, 
${\vec p_{1\parallel}} = {\vec p_{2\parallel}}$.
Equation (\ref{E}) then yields
\begin{equation}
p_{1\perp}^2 + \mu_{s1}^2 = p_{2\perp}^2 + 
\mu_{s2}^2 \equiv \varepsilon_\perp\ ,
\label{Eperp}
\end{equation}
defining a conserved quantity $\varepsilon_\perp$.  
We now see how to
translate results obtained in the $(1+1)$-dimensional treatment
of the previous section into the full $(3+1)$-dimensional 
setting appropriate here.  What was called $E$ in previous
sections is in fact $\varepsilon_\perp$.  $\Delta R$ depends on
$\varepsilon_\perp$, and the enhanced reflection zone is given by 
$\mu_{s2}<\varepsilon_\perp<\mu_{b2}$.  

Before the arrival of the wall, the $\phi$'s are in thermal equilibrium
with momenta distributed according to the Bose-Einstein distribution.
This equilibrium distribution defines a rest frame, 
the plasma frame, which is  different from the
rest frame of the wall.  
In order to compute ${\cal F}_B$, we need the flux density
of $\phi_i$ particles (equivalently, $\bar\phi_i$ 
antiparticles) incident upon the wall with a given momentum 
in the wall frame.
Throughout this section, we continue to assume that the mean free
path of particles is larger than the wall width, deferring the
discussion of the effects of the falseness of this assumption
to Section IV.  Hence, the incident flux density we require is
given simply by 
\begin{equation}
f_i(E,p_{i\perp}) = 
{3 p_{i\perp}/E \over e^{\gamma(E - v_w p_{i\perp})/T} - 1}, 
\label{Fi}
\end{equation} 
where $\gamma \equiv 1/\sqrt {1 - v_w^2}$.
The factor of 3 appears because there are $\phi$'s with
each of 3 different colors in thermal equilibrium. 
The argument of the exponential arises because,
as mentioned before, the particles are initially in thermal
equilibrium in the plasma frame, not in the wall frame.
To get $F_{s \to s}$, 
we must integrate $\Delta R \equiv R_{12} - 
R_{\overline{12}} = R_{12} - R_{21}$ against $f_1$ 
over the three-momentum of the incident $\phi_1$ and integrate 
$R_{21} - R_{\overline{21}} = R_{21} - R_{12} = - \Delta R$ 
against $f_2$ over the three-momentum of the 
incident $\phi_2$ and add up the results.  
Thus, we obtain
\begin{eqnarray}
F_{s \to s}&=&{1 \over \gamma} \Bigg[ 
\int_{0}^{2 \pi}\int_{0}^{\infty} {d \theta \, d p_{1\parallel}\, 
p_{1\parallel} \over (2 \pi)^2}       
\int_{p_{1 \perp {\rm min}}}^{\infty} {d p_{1\perp} \over 2 \pi}\,
\Delta R(p_{1\perp}) \,  f_1(E,p_{1\perp}) \Bigg]\nonumber\\ 
&-& {1 \over \gamma} \Bigg[\int_{0}^{2 \pi}\int_{0}^{\infty} 
{d \theta \, d p_{2\parallel}\, p_{2\parallel} \over (2 \pi)^2}       
\int_{p_{2 \perp {\rm min}}}^{\infty} {d p_{2\perp} \over 2 \pi}\,
\Delta R(p_{2\perp}) \,  f_2(E,p_{2\perp}) \Bigg] ,
\label{FSS}
\end{eqnarray}
where $p_{i \perp {\rm min}}$ is 
the $p_{i\perp}$ such
that $\varepsilon_\perp = \mu_{s2}$. Note that 
$F_{s\to s}$ is the flux seen in the plasma frame.
Transforming from the wall frame back to the plasma frame
yields the overall factor of $1/\gamma$ in (\ref{FSS}).
Upon performing the integration over $p_{i\parallel}$, we obtain
\begin{equation}
{\cal F}_B \simeq \frac{1}{3}F_{s \to s} = {T \over 4 \pi^2 \, \gamma^2} 
\int_{\mu_{s2}}^{\infty}
d \varepsilon_{\perp}\,\, \varepsilon_{\perp}\, \Delta R(\varepsilon_{\perp}) 
\ln\left[{1 - e^{-\gamma (\varepsilon_\perp - 
v_w p_{2\perp})/T} \over 1 - e^{-\gamma (\varepsilon_\perp - 
v_w p_{1\perp})/T}}\right],
\label{FSSp1L}
\end{equation}
where we have used (\ref{Eperp}) in the form
$\varepsilon_\perp d\varepsilon_\perp = 
p_{1\perp} d p_{1\perp}=
p_{2\perp}d p_{2\perp}$.

{}From (\ref{FSSp1L}), we deduce that $F_{s \to s} = 0$ for 
$v_w = 0$.  This is to be expected, since baryogenesis requires
out of equilibrium conditions, and hence requires $v_w\neq 0$.
(Note that one can derive an expression for $F_{b \to s}$ 
similar to that for $F_{s\to s}$ 
and show that $F_{b \to s} = 0$ when $v_w = 0$.) 
The vanishing of  ${\cal F}_B$ with $v_w$ can 
be more directly understood as follows.
We present the
argument in $1+1$
dimensions; the generalization to $3+1$ is trivial.
We have shown that $R_{12}-R_{\overline{12}}=-(R_{21}-R_{\overline{21}})$.
For $v_w=0$, the number of $\phi_1$ particles incident upon
the wall with an energy $E$ greater than 
$\mu_{s2}$ is equal to the number of incident
$\phi_2$ particles with the same energy.  (There are of course $\phi_1$ 
particles with $\mu_{s1}<E<\mu_{s2}$, but they do not yield
a reflection asymmetry.)  Therefore, for $v_w=0$, the
contribution to ${\cal F}_B$ due to incident $\phi_1$'s
is exactly cancelled by that due to incident $\phi_2$'s.
Now consider a moving wall, where $v_w\neq 0$.
$\phi_i$ particles incident upon the wall with 
a given energy $E$ in the 
wall frame have energy $(E- v_w p_{i\perp})$
in the plasma frame, in which
they are initially in a thermal equilibrium
distribution.  If $\mu_{s1}\neq \mu_{s2}$ 
then $p_{1\perp} \neq p_{2\perp}$
for a given $E$, and the number of $\phi_1$ and $\phi_2$ 
particles with incident
energy $E$ in the wall frame is {\it not} the same.  We see that
in order to upset the cancellation between the contribution to
${\cal F}_B$ due
to incident $\phi_1$'s and that due to $\phi_2$'s, we need 
both $v_w\neq 0$ and $\mu_{s1}\neq \mu_{s2}$.
The asymmetry in the reflection coefficients can only yield
an asymmetry in the baryon number flux if the wall is moving
and if the scalars are not degenerate in mass in the
symmetric phase.

As discussed in the previous section, 
$\Delta R$ 
depends on parameters in $M^2$ and on the wall width.
The flux ${\cal F}_B$ depends on these parameters as
well as on the wall velocity and temperature.
We will be interested in the regime in which $v_w$ and
$\Delta m/m$ are small compared to $1$.  This may seem
surprising, given that we have just argued that ${\cal F}_B$
is zero for $v_w\rightarrow 0$ or $\Delta m/m\rightarrow 0$.
The reason is that, as we have
seen in the previous section, $\Delta R$ 
is a decreasing function of $\Delta m$ and, as we will see in
the next section, the BAU is proportional to ${\cal F}_B/v_w^3$
for realistic wall velocities.
In order to gain intuition about (\ref{FSSp1L})
it is useful to pretend that $\Delta R$ is energy
independent for $\mu_{s2} < \varepsilon_\perp < \mu_{b2}$,
and then expand in $v_w$ and $\Delta m/m$ to
first order in both, obtaining
\begin{equation}
\frac{{\cal F}_B}{T^3}\sim\frac{1}{4\pi^2}\,v_w\,\frac{\Delta m}{m}\,
\left(\frac{m}{T}\right)^2\,\Delta R\,\int_{m/T}^{\mu_{b2}/T}\frac{y\,dy}
{\left[e^{y}-1\right]
\sqrt{y^2-(m/T)^2}}\ .
\label{expansion}
\end{equation}
We perform all our calculations using (\ref{FSSp1L}), not
the expansion (\ref{expansion}), but the expansion
is useful for understanding the qualitative dependence of
${\cal F}_B$ on the parameters.   

With all parameters as in our canonical example,
(\ref{FSSp1L}) yields
\begin{equation}
\frac{{\cal F}_B}{T^3}\sim 9\times 10^{-6}\ ,
\label{prefb}
\end{equation}
and with $w=5/T$ instead of $25/T$, we obtain a result which 
is a factor of two larger.
We have verified 
that ${\cal F}_B$ is linear in $v_w$ to within a few percent
for $v_w<0.6$.  In the regime in which ${\cal F}_B$ is linear in
$\Delta \beta$, $\Delta \varphi$, and $v_w$, (\ref{prefb}) becomes
\begin{equation}
\frac{{\cal F}_B}{T^3}\sim 2 \times 10^{-4}\,v_w
\,\Delta\beta\,\Delta\varphi\ .
\end{equation}
\begin{figure}
\centerline{
\epsfysize=3.5in
\epsfbox{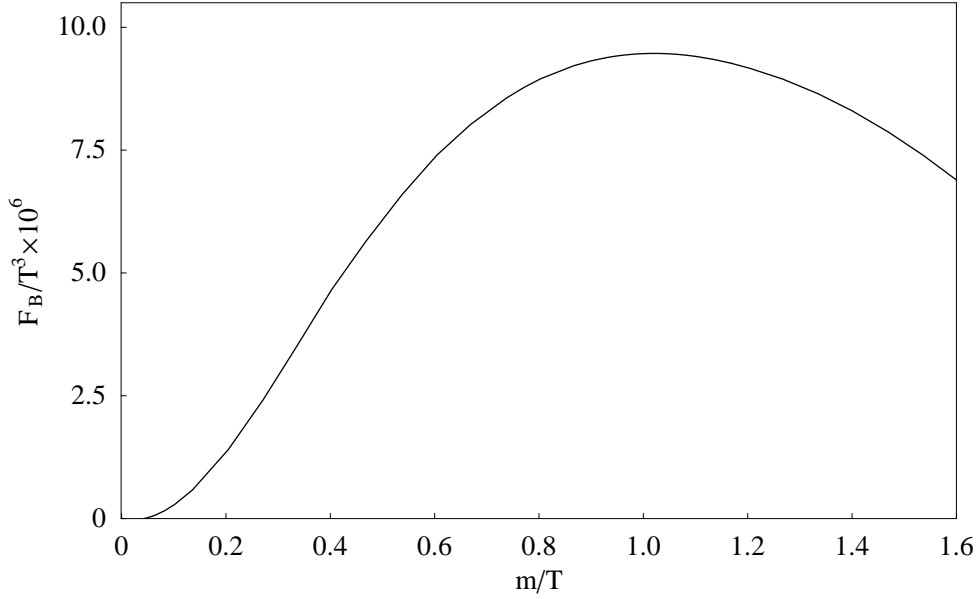}}
\caption{Temperature dependence of
the baryon number flux.  In this plot, $T$ varies and
$w=25/T$ for all $T$.  We keep 
all parameters except $T$ and $w$ fixed.
We plot ${\cal F}_B/T^3$ vs. $m/T$ to facilitate
comparison with (\ref{expansion}).
}
\label{fig5}
\end{figure}

Turning now to the temperature dependence,  
in Figure 5, we plot ${\cal F}_B/T^3$ vs. $m/T$, varying
$T$ and keeping all parameters in the mass matrix fixed.  
Since $\Delta R$ does not depend on $T$, we can partially understand
this plot by noting that in  
(\ref{expansion}) ${\cal F}_B/T^3 \sim (m/T)^2$ at small $m/T$
and $\sim \exp(-m/T)$ at large $m/T$. 
This does not completely describe the Figure, however, because
as we vary $T$, we have kept $w=25/T$; this means that $w$
changes with respect to the parameters in the mass matrix.
We conclude from Figure 5 that
the BAU generated by the scalar baryon number transport
mechanism is largest for $m\sim T$.
\begin{figure}
\centerline{
\epsfysize=3.5in
\epsfbox{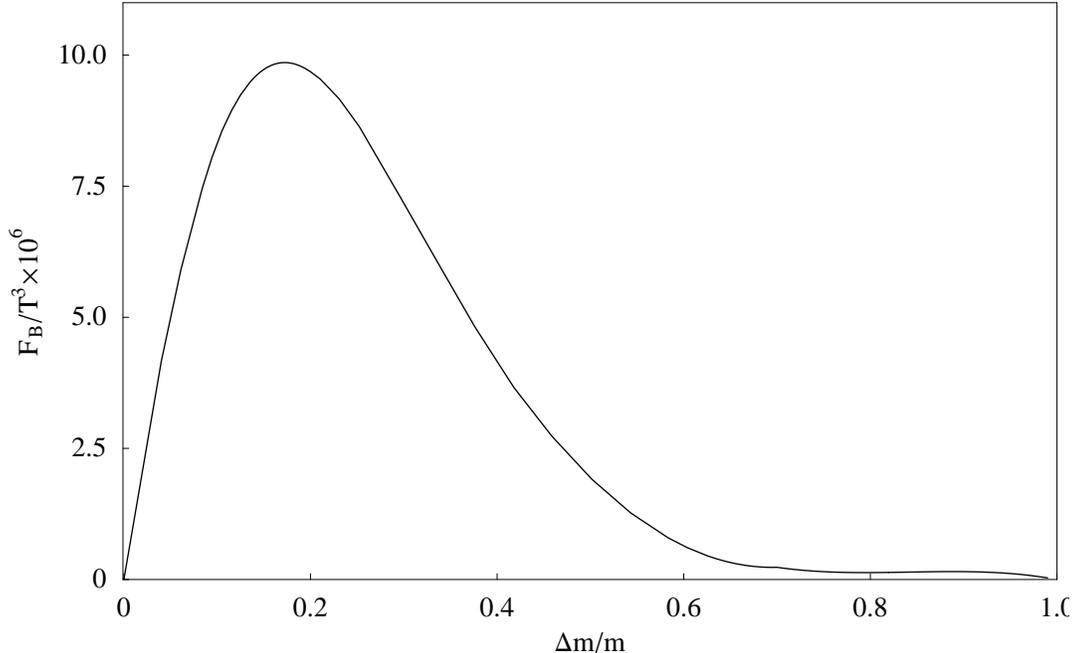}}
\caption{Dependence of the baryon number flux on $\Delta m/m$, with
$\Delta m$ varying and all other parameters fixed.
}
\label{fig6}
\end{figure}

In Figure 6, we plot ${\cal F}_B/T^3$ vs. $\Delta m/m$.
It is linear in $\Delta m/m$ for small $\Delta m/m$
and falls at large $\Delta m/m$ because, as we 
noted in the previous section, $\Delta R$ decreases with increasing
$\Delta m/m$.  We see that using $\Delta m/m=0.2$ as in
our canonical example yields a reasonable estimate 
for $\Delta R$ over the range
$0.1 < \Delta m/m < 0.3$, but for $\Delta m/m$ outside this
range, the BAU is suppressed. 
For the scalar baryon transport mechanism to be efficient,
we need scalars with symmetric phase masses of ${\cal O}(T_c)$
that differ by 10-30\%.

\section{Estimating the Baryon Number of the Universe}

In the preceding sections, we have shown how to calculate
the baryon number flux ${\cal F}_B$ carried by $\phi$ particles
that is injected into the symmetric phase by the motion of
the bubble wall.  To this point, we have described the quantitative
solution of a well-posed problem.  Given a mass matrix, 
a critical temperature, a wall
profile, a wall velocity, and making the assumption that
the $\phi$ mean free path is long compared to the wall width,
a quantitative calculation of ${\cal F}_B$ is attainable.
In this section,
we sketch a qualitative estimate of the cosmological
baryon to entropy ratio, $n_B/s$, that results from the
flux ${\cal F}_B$.  Our treatment is admittedly crude and
can be improved, for example along the lines of that
of Huet and Nelson \cite{HN}, but we leave this for future work.
We organize the estimate of the final result as follows.
First, we estimate the mean free path $l$ and the suppression
of ${\cal F_B}$ that results from the finiteness of $l/w$. Then,
we estimate the scalar baryon number density (baryon number
in the form of $\phi$'s) that results from ${\cal F}_B$.
This in turn leads to a quark baryon number density which
biases the electroweak sphaleron processes acting in the
symmetric phase, 
resulting in a net baryon asymmetry of 
the universe.

Before the wall arrives, the $\phi$'s in the symmetric
phase are diffusing about 
with a mean free path in the $x$ direction which we call $l$
and a mean velocity in the $x$ direction between scatterings 
which we call $v_\phi$.  The one-dimensional diffusion constant
is defined as
\begin{equation}
D = l\,v_\phi \ .
\label{Ddef}
\end{equation}
The mean one-dimensional velocity is $v_\phi\sim 0.7$ for
particles with $m=T$.  Joyce, Prokopec, and Turok\cite{JPT}
have estimated that in the symmetric phase, the diffusion constant
for relativistic strongly interacting particles is
$D\sim 6/T$.  Most of the scatterings contributing
to $D$ are interactions with gluons in the plasma,
and $D$ is inversely proportional to the relevant
interaction cross-section
which decreases with increasing $m$.
This suggests that for $m=T$, the diffusion
constant is somewhat greater than $6/T$, but we leave
a complete calculation for future work.  In this paper,
we simply use the conservative estimate
\begin{equation}
l\sim \frac{10}{T}\ .
\label{ldef}
\end{equation}
Note that an increase in $m/T$ leads to an increase in $l$, and
potentially to a larger final result.  Of course, in increasing
$m/T$, one pays a penalty in the exponential suppression
in (\ref{expansion}), and we will see below that making
the $\phi$'s nonrelativistic exacts 
other costs as well.  

We now give a crude estimate of 
the suppression due to the finiteness of $l/w$.
As noted in Section II, estimates for $w$ range from $10/T$ to $100/T$.
In our canonical example, we use $w=25/T$, yielding $l/w\sim 0.4$.
The essence of the effect is
that when $\phi$ particles reach $x_0$, the point in the
wall at which one mode is totally reflected, they have
only been travelling (and mixing) freely for a distance 
of order $l$.
To incorporate this, we redo the calculation of $\Delta R$
as follows. For each incident energy, we find $x_0$ and choose  
as incident states the mass eigenstates a distance $l$
to the left of $x_0$.  These then propagate only a distance 
$l$ before reflecting, and so experience less $CP$ violating
mixing than in the case where $l$ is infinite.  The result is
a suppression in $\Delta R$ at each energy. We in fact find that
this suppression is rather energy dependent, being larger for
energies close to $\mu_{b2}$.  In evaluating ${\cal F}_B$,
therefore, we must re-evaluate the integral (\ref{FSSp1L}).
\begin{figure}
\centerline{
\epsfysize=3.5in
\epsfbox{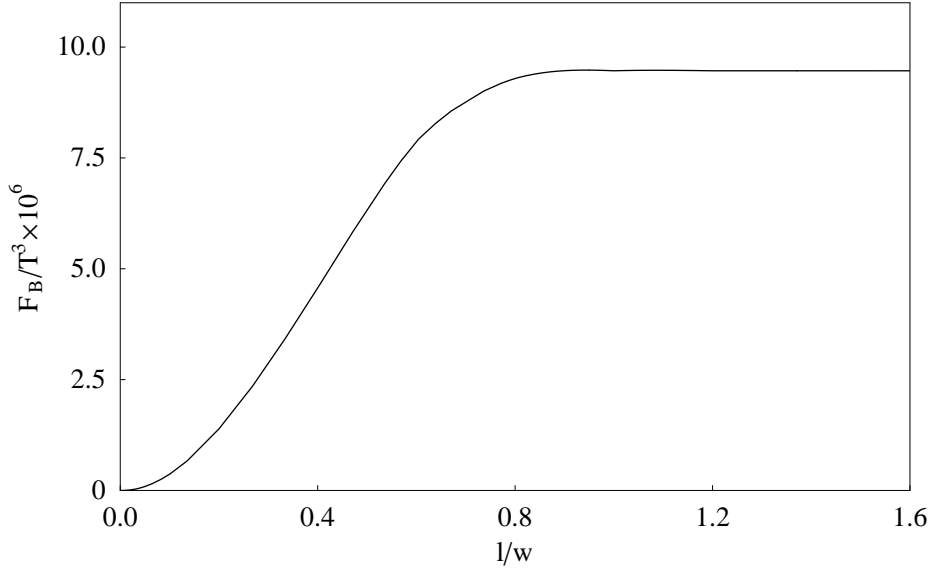}}
\caption{Dependence of the baryon number flux on $l/w$, with
$l$ varying and all other parameters fixed.
}
\label{fig7}
\end{figure}
The result is shown in Figure 7, in which we plot 
${\cal F}_B/T^3$ vs. $l/w$. We see
that for $l/w=0.4$, the flux ${\cal F}_B$ is suppressed by
about a factor of $2$ relative to that for infinite $l$.
We have also verified that the reflection asymmetries at a given
energy are largely
insensitive to the form of the potential beyond the corresponding $x_0$.
Although our method of including the effects of a finite
mean free path is certainly not the final word, it
should give a reasonable estimate of the magnitude of
the suppression.  Note in particular that the integrand in
(\ref{FSSp1L}) is largest for lower energies within the
enhanced reflection zone. For these energies, $x_0$ is
on the symmetric phase side of the wall, and the
dependence on $l/w$ is not severe. This qualitative explanation
is consistent with our result that ${\cal F}_B$ is only 
suppressed by a factor of 2 for $l/w=0.4$.

If $l$ is larger than we have estimated,
as for example would be the case for more non-relativistic
$\phi$'s for which $D$ is larger and $v_\phi$ is smaller,
then the final result could be larger than that which
we estimate by up to a factor of $2$.  If, on the other hand,
$w$ is larger than $25/T$, the final result will be suppressed
both by the effect of $wm$ displayed in Figure 3 and by the
effect of $l/w$ displayed in Figure 7. Of course, $w<25/T$ 
would correspondingly enhance the final result, for example by a
factor of 4 for $w=l=10/T$.  
In the estimates that follow,
we take ${\cal F_B}/T^3\sim 5\times 10^{-6}$, as appropriate
for $m=A=\mu=T=100 {\rm GeV}$, $\Delta m/m = 0.2$, 
$w=25/T$, $l/w=0.4$, $v_w=0.1$ and $\Delta \beta\Delta\varphi=0.5$. 
In the regime in which ${\cal F_B}$ is linear in $v_w$, $\Delta
\beta$, and $\Delta \varphi$,
we can write it as 
\begin{equation}
\frac{{\cal F}_B}{T^3} 
\sim  1\times 10^{-4} \,v_w\,\Delta \beta\,\Delta\varphi\ .
\label{finalFB}
\end{equation}
The dependence on the other parameters is more complicated, as we
have discussed and illustrated above.

Next, we give an estimate of the 
baryon number density carried by $\phi$ particles in
the region in front of the bubble wall.  A $\phi$
particle emerging into the symmetric phase from the
wall begins to diffuse, and the mean distance
such particles have travelled from the wall a time $t$ after
being reflected is  $x \sim \sqrt{2 l v_\phi t}$.  In the same time
$t$, the wall itself has moved a distance $v_w t$.  
Defining $\tau$ as the time that a reflected $\phi$ particle spends
in the symmetric phase before the wall overtakes
it, we find that on average,
\begin{equation}
\tau \sim \frac{2 l v_\phi}{v_w^2}\ \label{taudef}.
\end{equation}
Note that this means that on average a reflected $\phi$ particle
undergoes 
\begin{equation}
N_{\rm scatt} \sim \frac{\tau}{l/v_\phi} 
\sim 2\,\left(\frac{v_\phi}{v_w}\right)^2
\label{numberofscatt}
\end{equation}
scatterings during the time $\tau$ it spends in the broken phase.
Our treatment is only consistent for  $N_{\rm scatt}> 1$, and our
final result is largest for large $N_{\rm scatt}$.  
As mentioned above, although non-relativistic $\phi$'s have
benefits, they also have costs, and we see one here.

We denote the mean separation between the diffusing particle and the
oncoming wall during the time the particle is in the
broken phase by $\Delta x$.  This quantity 
will cancel in the
final result.
Over a range of $x$ in front of the wall given approximately by $\Delta x$,
there is a net
baryon number density carried by $\phi$ particles
given by 
\begin{equation}
n_B^\phi \sim {\cal F}_B\, \frac{\tau}{\Delta x} \ .
\label{nbphi}
\end{equation}
We arrive at this estimate by noting that ${\cal F}_B$
is the baryon number injected into the symmetric phase
per unit wall area per unit time and that 
at any given time, the $\phi$'s reflected
in the previous $\tau$ are in a region of order 
$\Delta x$ ahead of the wall. 
Thus, we conclude that every point in the universe
experiences a baryon density in $\phi$ particles given
by (\ref{nbphi}) for a time  
\begin{equation}
t^* \sim \Delta x /v_w 
\label{timeforfun}
\end{equation}
while in the symmetric phase.  
By this point, it should
be becoming clear that $n_B/s$ will turn out to be
largest for small wall velocities. The authors of
Refs. \cite{linde,mp} discuss wall velocities ranging
from $0.02$ to $0.4$, and other authors have considered velocities
as high as $0.9$. The mechanism we are 
proposing will be most effective at the lower end of this
range, and we have been using $v_w\sim 0.1$ in our canonical example.  
Note also that we are assuming that $n_B^\phi/T^3$ is 
small, and hence we are not including the effect of the
baryon number asymmetry in the distribution
functions used in the calculation of ${\cal F}_B$ in the previous section.

The mechanism that generates the baryon number flux
carried by $\phi$ particles into the symmetric phase does not involve any 
baryon number violation. 
Therefore, the baryon number associated with $n^\phi_B$  
must be exactly
cancelled by a baryon number of opposite sign behind the wall.  
Note that the $\phi$ particles, being scalars, are not 
produced anomalously in electroweak sphaleron processes,
and therefore cannot bias such processes. Thus, if this
were the end of the story, we would have made no progress.
However, $\phi$ particles
can be converted into quarks.  This is a model-independent statement,
equivalent to the statement that the $\phi$'s have
baryon number $1/3$. The rate for $\phi$--quark conversion
is, however, model dependent.
In the supersymmetric case we are using as an example,
if the gluino mass $M_{\tilde g}$ is less than $\mu_{s2}$, the symmetric phase
mass of the heavier squark, 
then the squarks can decay into gluinos and
quarks, since the quarks are massless in the symmetric phase.
We will not assume that the gluinos are this light, however.
The first process we consider is scattering off a gluino
in the thermal bath:
squark $+$ gluino $\rightarrow$ quark $+$ gluon.  The
rate for this process is suppressed relative to that
for squark-gluon scattering by $\exp(-M_{\tilde g}/T)$
due to the paucity of gluinos in the plasma.  It is also
suppressed by $(T/M_{\tilde g})^2$ in the cross-section.
The second process we consider is gluon $+$ squark $\rightarrow$
quark $+$ virtual gluino, where the gluino becomes quark $+$ anti-squark
or anti-quark $+$ squark.   This rate is suppressed relative
to ordinary quark-gluon scattering by of order $\alpha_s (T/M_{\tilde g})^2$,
and by three-body phase space.
Defining $1/b$ as the fraction of scatterings incurred by
a $\phi$ particle in the symmetric phase that turn the $\phi$
into a quark, we estimate that $b\sim 200$
for $M_{\tilde g}\sim 300~{\rm GeV}$, noting again that this
estimate is quite model dependent.
We can now
estimate that the baryon number density carried by quarks 
in the region $\Delta x$ in front of the wall is
\begin{equation}
n^q_B \sim n^\phi_B \,\frac{N_{\rm scatt}}{b}\sim 
n^\phi_B\,\frac{2\,v_\phi^2}{b \,v_w^2} \ ,
\label{nqb}
\end{equation}
where we have used (\ref{numberofscatt}).  This estimate is only
valid for $N_{\rm scatt}/b < 1$; if $N_{\rm scatt}/b > 1$ then 
the squark--quark conversion reactions have time to establish 
chemical equilibrium, and an analysis in terms of chemical
potentials should be used.  For $N_{\rm scatt}/b\sim 1$,
which is relevant for $M_{\tilde g}\sim 300~{\rm GeV}$ and $v_w\sim 0.1$, 
the average $\phi$ is converted to a quark at some point during
its $N_{\rm scatt}$ scatterings, which leads us to estimate that
the baryon number density in quarks is
\begin{equation}
n^q_B\sim \frac{n^\phi_B}{2} \sim {\cal F_B} \frac{\tau}{2\Delta x}\ ,
\label{nqb2}
\end{equation}
where we have used (\ref{nbphi}). 
The density $n^q_B$ {\it does} bias electroweak sphaleron processes.
Note that a baryon number density in front of the wall cannot
be affected by non-perturbative QCD processes. This 
is in contrast to what happens in many other mechanisms.
For example, if an axial baryon number density (more
left handed quarks than right handed ones; no net excess
of baryons) is generated, this can bias electroweak sphaleron
processes only if it is not first wiped out by QCD processes.

The rate per unit volume of baryon number violating sphaleron
processes in the symmetric phase is conventionally written as
\begin{equation}
\Gamma_s \sim \kappa \alpha_W^4 T^4 \ .
\label{kappadef}
\end{equation}
Recent work\cite{ASY} shows that $\kappa$ is parametrically
${\cal O}(\alpha_W)$.  The most complete numerical simulations done
to date\cite{MHM} suggest $\kappa=(29 \pm 6) \alpha_W$.
In thermal equilibrium, no net baryon asymmetry is generated
by these processes, since sphaleron and ``anti-sphaleron'' processes
occur at the same rate.  However, in a region with a nonzero
$n_B^q$, the sphaleron processes tend to reduce the 
baryon number density. In particular\cite{review}
\begin{equation}
\dot n_B \sim - \,6\, n_B^q \,\frac{\Gamma_s}{T^3}\ .
\label{gammadot}
\end{equation}
Assuming that 
$\Gamma_s t^* /T^3 < 1$, which is
certainly the case for $v_w>0.005$, then the net 
change in $n_B$ due to anomalous electroweak processes is
\begin{equation}
\Delta n_B = -\, 6\, n_B^q\, \frac{\Gamma_s}{T^3} t^* \ .
\label{deltanb}
\end{equation}
Long after the electroweak phase
transition, after the universe has re-homogenized, the
remaining excess baryon number density will be given by $\Delta n_B$.
Note that in order to end up with a baryon density in the universe
today, the phase in the mass matrix $M^2$ must be such that 
${\cal F}_B$ is negative. A net flux of anti-baryons is injected
into the symmetric phase, compensated by a baryon flux from the
wall into the broken phase. Some fraction of the anti-baryon
excess in front of the wall is wiped out by sphaleron processes and
after the entire universe is swept by the broken phase, a positive
baryon number asymmetry persists at temperatures below $T_c$. 

Using (\ref{deltanb}), (\ref{kappadef}), (\ref{nqb2}), (\ref{timeforfun})
and (\ref{taudef}), and noting
that the cosmological entropy density at the time of the
electroweak phase transition is $s=(2\pi^2 g^*/45)T^3\sim 55 T^3$
with $g^*$ the number of degrees of freedom in equilibrium, 
we estimate
that the mechanism we have presented yields a baryon to
entropy ratio\footnote{Although our final expression has powers of
$v_w$ in the denominator, the derivation relies on (\ref{deltanb}),
which is only valid for $\Gamma_s t^* /T^3 < 1$, and the BAU does not
in fact diverge for $v_w \rightarrow 0$.}
\begin{equation}
\frac{n_B}{s} \sim \frac{6}{55}\, \kappa \alpha_W^4\, 
\frac{l\,T\,v_\phi}{v_w^3}\,\frac{{\cal F}_B}{T^3}\ .
\label{firstfinal}
\end{equation}
Inserting the expression (\ref{finalFB}) for the baryon number flux
${\cal F}_B$, valid in the regime in which ${\cal F}_B$ is linear
in $v_w$, $\Delta \beta$ and $\Delta
\varphi$,
and using $v_\phi \sim 0.7$ and $l\sim 10/T$, we obtain
\begin{equation}
\frac{n_B}{s} \sim 1\times 10^{-10}\, \frac{\kappa}{v_w^2} \,
\Delta\beta\,\Delta\varphi\,\ .
\label{finalresult}
\end{equation}
As we have discussed, this result is obtained for baryon
number carrying scalars with symmetric phase masses of
${\cal O}(T_c)$ that are non-degenerate by $\Delta m/m \sim 0.1-0.3$.
The result (\ref{finalresult}) is sensitive to the wall
width $w$.  If we optimistically use $w=10/T$ instead
of $w=25/T$, the BAU increases by a factor of 
$4$ relative to that of (\ref{finalresult}).   

The asymmetry (\ref{finalresult})
is at an interesting level.  For example, if $v_w \sim 0.1$,
a BAU consistent with cosmological observation is obtained for
\begin{equation}
\Delta \beta \Delta\varphi > 0.003,
\label{bound}
\end{equation}
where we have used $\kappa \sim 29 \alpha_W \sim 1$\cite{MHM}.
Taking $v_w=0.1$ as we have done
is reasonable, but wall velocities as low as $v_w\sim 0.02$
are possible \cite{linde,HN}.  Therefore, making the most optimistic 
choice for $v_w$, we find that the scalar 
baryon number transport mechanism can yield a BAU consistent with
cosmological observation for $\Delta\beta\Delta\varphi>1\times 10^{-4}$,
or even $\Delta\beta\Delta\varphi>3\times 10^{-5}$ if $w=10/T$.  
Making more reasonable choices for $w$ and $v_w$
yields (\ref{bound}).  We see that even if we use the conservative
estimate $\Delta\beta\sim 0.01-0.03$\cite{macarena,quiros}, no strong
constraints need be imposed on $\Delta\varphi$ in order for our
implementation of the scalar baryon number transport mechanism using
top squarks to yield a BAU consistent with cosmological observation.

We have demonstrated that the scalar baryon number transport
mechanism can yield a cosmologically interesting BAU, and
have done so via a supersymmetric  implementation of the mechanism.
It is therefore interesting to compare our result to those
of other authors who have studied electroweak baryogenesis
in supersymmetric theories.  Most recent 
treatments\cite{susynowashout,macarena,aoki,worah} have
included one stop with a light broken phase mass, in order
to have a strongly first order phase transition so that
the BAU which is generated is preserved\cite{susynowashout},
but they have used symmetric phase stop masses which are 
much larger than $T_c$.  In this regime, our mechanism is
not effective.  These treatments have used 
particles other than stops and have used the charge transport
mechanism. Given the new results of Moore\cite{moore}, they
have likely underestimated the effects of strong interaction
processes which wash out axial baryon number.  Nevertheless, it
seems likely that for masses
such that the scalar baryon number transport mechanism is
efficient, the contribution from charge transport involving particles
other than stops
is comparable to that which we find.\footnote{All the mechanisms we
discuss in this paper are ``non-local'', in the sense that
the relevant CP violation occurs at the bubble walls, while
the relevant B violation occurs away from the bubble walls
in the symmetric phase.  Local mechanisms, in which 
CP violation acts directly to bias the gauge and Higgs
dynamics of sphaleron processes, have also been 
considered\cite{TZ,DHSS,CKN3,LRT}
and also make a contribution to the BAU. The work
of Ref. \cite{LRT} suggests that for the thin wall case
of interest in this paper, the local contribution to the
BAU is likely small.}

The one treatment other than ours that includes reasonably
small symmetric phase stop masses is the
work of Huet and Nelson\cite{HN},
and we now attempt a more quantitative comparison with their results.
For one of their sets of parameters,\footnote{Note that 
a completely quantitative
comparison between our results and those of Huet and Nelson 
is actually not possible,
because they choose to neglect those diagonal 
terms in $M^2$ of (\ref{squarkmass}) proportional to $m_Z^2$ and $m_t^2$.
This may be justifiable for $m=150~{\rm GeV}$, and we therefore
compare our results with those they obtain using this parameter set.
For their other set of parameters, which has
$m=60~{\rm GeV}$, $A=\mu=50~{\rm GeV}$, their $M^2$ has one negative eigenvalue
in the broken phase, rendering comparison to results they obtain
with these parameters difficult.}
namely
$m=150~{\rm GeV}$, $A=\mu=50~{\rm GeV}$ and $T_c=60~{\rm GeV}$
they find a contribution to the BAU due to stops using the 
charge transport mechanism
given by  
$n_B/s \sim 2.3\times 10^{-10} 
v_w (\kappa/\kappa') \Delta\beta\Delta\varphi$.  Just as $\kappa$
parametrizes the rate for electroweak baryon number
violating processes, $\kappa'$ parametrizes the rate
for strong axial baryon number violating processes.
The factor $(\kappa/\kappa')$ is conventionally taken to be $\sim 1$, but
the work of Moore\cite{moore} suggests that it is in fact smaller.
For $v_w\sim 0.1$, the BAU
(\ref{finalresult}) generated by the scalar baryon number
transport mechanism is a factor of $400\kappa'$ 
larger than that generated by 
charge transport involving stops.   Although the mass
matrix used in Ref. \cite{HN} is not given by (\ref{squarkmass}),
it seems plausible that 
when the scalar baryon number transport mechanism is efficient, namely 
for $m\sim T$, $\Delta m/m \sim 0.1-0.3$, it yields
the dominant stop contribution to the BAU.  Huet and Nelson
also consider the contribution to the BAU due
to charge transport involving particles other than stops and find 
$n_B/s \sim 6.5\times 10^{-9} 
v_w (\kappa/\kappa') \Delta\beta\sin\varphi_B$.  This is somewhat smaller
than (\ref{finalresult}), but only by a factor of $15\kappa'$
if $\varphi_B\sim\Delta\varphi$.  Note, however, that
in some models\cite{falk} $\varphi_B$ is suppressed
while $\Delta\varphi$ is not.
Nevertheless, a
complete treatment of the BAU should include the contribution due to
charge transport involving 
particles other than stops.
Although perhaps not the whole story, scalar baryon number transport 
yields the dominant contribution to the BAU due to stops, and
can explain the cosmologically observed value.

\section{Open Questions and Model Implementations}

We have given our quantitative conclusions in the final four paragraphs of
the previous section; the present section is devoted to unresolved questions
and to a discussion of possible implementations of
the scalar baryon number
transport mechanism.  We have organized
our presentation of the scalar baryon number transport mechanism
in such a way that all the parts of the treatment requiring technical
improvement
were deferred to Section IV, in which we restricted ourselves
to making estimates.  
For example, further work is certainly
called for in the calculation of the diffusion constant $D$ 
for semi-relativistic strongly interacting particles.  In addition,
our treatment of the suppression due to finite mean
free paths can be improved.  The reader can surely find other
ways to improve the arguments of Section IV.
Moving beyond the
technical, we noted 
at the end
of Section IV that a more complete treatment should include 
the transport of charges other than baryon number carried by
the scalars of interest.  Also, we have neglected thermal
contributions to particle masses.  Since they are of order
couplings times $T_c$, and since the scalars we discuss have
masses of order $T_c$
in the symmetric phase and somewhat higher in the broken phase,
neglecting thermal masses seems reasonable in this  exploratory
treatment of the scalar baryon number transport mechanism,
and we have left their inclusion to future work.

The crucial observation that makes the scalar baryon
number transport mechanism possible is the existence
of a broad enhanced reflection zone, a
range of incident
energies in which reflection coefficients and their
CP violating asymmetries are large.  This arises
when there are a different number of propagating modes
at a given energy on the two sides of the wall.
A broad enhanced reflection zone may arise in contexts other than that
which we have considered, for example with scalars that do not carry
baryon number.  This suggests that insights gained from this work may
have wider application.

As noted in the introduction, our main goal in this paper
has been to present the scalar baryon number transport
mechanism, not to address model building issues.
We have chosen to work within a supersymmetric scenario.
Within this context, we now discuss some lessons for
future model building efforts.  It has
already been realized\cite{susynowashout} that it is
desirable for one stop to have a zero temperature mass less than 
the top mass, because this assists in making the electroweak phase
transition more strongly first order.  We now see
that it is also advantageous to have symmetric phase masses
that are of order $T_c \sim 100 ~{\rm GeV}$, 
and that differ by 10-30\%.
It will be interesting to look for supersymmetric models
satisfying this criterion.  As noted in Section II, non-degeneracy
of the appropriate magnitude
arises in some models\cite{falk} due to renormalization group running 
down
from a high energy scale at which the stops are degenerate.

We have called our mechanism scalar baryon number transport rather
than stop baryon number transport deliberately.  The essential
features of our mechanism can be implemented in other extensions
of the standard model involving baryon number carrying scalars,
although such extensions are perhaps not as well motivated 
as the supersymmetric scenario.
The recent HERA anomaly\cite{HERA} may hint at the existence
of first generation scalar leptoquarks of 
zero temperature mass $\sim 200~{\rm GeV}$.
(See, for example, the treatment of Babu {\it et al.}\cite{babu}.) 
A scalar leptoquark is a particle with a Yukawa coupling 
to a quark and a lepton, which therefore carries both
lepton and baryon number.
Suppose that there are 3 generations of leptoquarks diagonally 
coupled to the 3 generations of quarks and leptons by their
Yukawa couplings.  
In a model with two Higgs doublets $H_1$ and $H_2$, 
couplings of the form $H_i H_j \phi_\alpha \phi_\beta$,
where $i,j=1,2$ and $\alpha,\beta=1,2,3$, can contribute
to the masses of the leptoquarks in the
broken phase and provide CP violating mixing.
If the leptoquark masses receive other contributions that 
are nonzero in the symmetric phase, then a mass matrix
of the required form can arise.   (In fact, if contributions
beyond tree-level are included, CP violating mixing terms
can arise even in a theory with a single Higgs field.)
The simplest possibility, namely mixing between 
first and second generation leptoquarks, is tightly
constrained by bounds arising from the
non-observation of flavor changing neutral currents\cite{babu},
and probably cannot yield large enough off-diagonal
terms in the mass matrix to be of interest.
However, such bounds are absent or much weaker for $\phi_2-\phi_3$
or $\phi_1-\phi_3$ mixing.  Implementing the scalar 
baryon number transport mechanism using leptoquarks is therefore
possible. It requires symmetric phase masses of ${\cal O}(T_c)$,
but since the zero temperature masses receive 
additional contributions proportional to the Higgs vacuum
expectation values, these can still be $\sim 200~{\rm GeV}$
or higher.  It is possible to construct leptoquark theories
that are consistent with experiment in which the
scalar baryon number transport mechanism generates a BAU 
consistent with observation.  We postpone further investigation,
in particular attempts at using the BAU to constrain parameters of
such theories, until such a time as the experimental evidence
becomes more compelling.

Let us hope nature is such that the stop (or leptoquark) spectrum is soon within
reach of experiment, enabling us to discover
whether the mass matrix is such
that the observed cosmological baryon asymmetry can be
due to scalar baryon number transport.
The mechanism is efficient
if there are two scalars with symmetric phase masses
of order $T_c$ that differ by 10-30\%, and if the bubble
walls during the electroweak phase transition are sufficiently slow
and sufficiently thin, as we have discussed in our conclusions
presented in the previous section.
Under such circumstances, the baryon number flux
produced by reflection of scalars with incident energies in 
the enhanced reflection zone can easily lead to a 
baryon asymmetry of the universe consistent with cosmological
observation.

\acknowledgments

It is a pleasure to thank John Preskill and Mark Wise
for many helpful suggestions.  We thank Martin Gremm for his careful
reading of the manuscript.  We have had useful discussions with
Greg Anderson, Sergey Cherkis, Edward Farhi, Patrick Huet, Anton Kapustin,
Zoltan Ligeti, Arthur Lue, Lisa Randall, Dam Son, Iain Stewart and
Mark Trodden.  This work was supported in part by
the Department of Energy under Grant No. DE-FG03-92-ER40701, and the
work of K.\ R.\ was supported in part by the Sherman
Fairchild Foundation.


\appendix
\section{Calculation of the Reflection Coefficients}

In this Appendix, we describe a method for numerically 
evaluating the reflection
coefficients used in Section II.  
We want to find solutions to the time-independent two field
Klein-Gordon equation (\ref{kg}).  Solutions to these
second order linear ordinary differential equations
are uniquely determined by specifying four boundary conditions
on the fields and/or their first derivatives.
Along with the linearity of the differential equations,
this implies that
the solutions form a linear vector space of complex dimension four.

As discussed in Section II, because particles incident from
the broken phase do not yield significant asymmetries, we need only
consider the problem of calculating the reflection coefficients
for  $\phi$'s incident from the symmetric phase.
In order to calculate $R_{12}$, for instance,
we must find a solution that 
satisfies the following conditions: At large
negative $x$ in the symmetric phase, $\phi_1$ has a right-moving
({\it i.e.} incident)
component with unit amplitude and $\phi_2$ has no right-moving
component.  There is no restriction
on the left-moving plane waves in the
symmetric phase --- their amplitudes determine
the reflection coefficients.
In the
enhanced reflection zone, the solutions in the broken
phase have one propagating mode, which must 
be purely right-moving,
and one non-propagating mode, which must decay
(rather than grow) exponentially
for $x\rightarrow\infty$.

In the relevant solutions, the propagating modes in
the broken phase must be purely right-moving ({\it i.e.} outgoing),
and the non-propagating modes must be exponentially decaying.
These
two broken phase boundary conditions, one for each mode, restrict the
space of relevant solutions to a two-dimensional subspace of the
complete four-dimensional solution space.  
In the basis of interest to us, the
two solutions spanning this subspace correspond to an incident
symmetric phase 
$\phi_1$ with no incident $\phi_2$ and to an incident symmetric phase
$\phi_2$ with
no incident $\phi_1$.  
To find these basis solutions directly requires imposing boundary
conditions in the symmetric phase. 
Imposing two boundary conditions at large positive $x$ and two at
large negative $x$ yields a more time consuming numerical
task than imposing four boundary conditions at one point.
Instead, we proceed as follows.
We first find two linearly independent solutions
satisfying the broken phase boundary conditions, but
not the symmetric phase boundary conditions.  We find each
solution 
by imposing four boundary conditions at one point in the
broken phase and using
the Runge-Kutta algorithm built into {\it Mathematica}\cite{Wolfram}.
These two solutions form a basis for the subspace
of solutions satisfying the broken phase boundary conditions,
and we find the basis of interest, namely the solutions
satisfying the symmetric phase boundary conditions, by
taking linear combinations.

The boundary conditions described above should in general
be imposed at spatial infinity.  
We find solutions satisfying boundary
conditions at finite $x_+$ in the broken phase and at finite $x_-$
in the symmetric phase.
Because we have chosen our 
our profile function $p(x)$ of (\ref{profile}) such that
the mass matrix does not vary for $x<-w/2$ and for 
$x>+w/2$, we can set $x_+=w/2$ and $x_-=-w/2$ without
loss of accuracy.
For a different
profile function, for instance one with exponential tails, one
would have to choose $x_-$ and $x_+$ far enough out on the tails
to achieve the desired accuracy.   

In Section IV, we discuss a method for obtaining a crude
estimate of the effects of a finite mean free path.
Instead of imposing boundary conditions at 
$x_-\rightarrow -\infty$ (equivalently for our
profile function, $x_-=-w/2$), we impose them
at $x_-=x_0-l$. Here, $x_0$ is the point where
one mode is totally reflected and is found
by setting one of the eigenvalues of the
mass-squared matrix equal to $E^2$ and $l$ is the mean free path.
In order to implement this calculation,
in the formalism we present below we keep
$x_-$ a free parameter.

We begin by finding two
solutions, $\mbox{solution}_\alpha$ and $\mbox{solution}_\beta$, 
by matching at $x_+$ to the following right-moving solutions:
\begin{eqnarray}
\mbox{$\mbox{solution}_\alpha$:  } &&
\vec{\phi}(x) = A e^{-i p_{b1} x} \vec{u}_{b1} \nonumber\\
\mbox{$\mbox{solution}_\beta$:  } &&
\vec{\phi}(x) = B e^{-i p_{b2} x} \vec{u}_{b2} 
\label{asymptoticsol}
\end{eqnarray}
where $p_{b1}=\sqrt{E^2-\mu_{b1}^2}$ and
$p_{b2}=\sqrt{E^2-\mu_{b2}^2}$
are the momenta of the normal modes in the broken phase at $x=x_+$;
$\mu_{b1}$ and $\mu_{b2}$ are the masses of these normal modes, defined
as the square roots of the eigenvalues of the mass-squared matrix in the
broken phase; and $\vec{u}_{b1}$ and $\vec{u}_{b2}$, the eigenvectors
of the broken phase mass-squared matrix, define the
broken phase
normal modes in the symmetric phase
$\vec{\phi}=(\phi_1, \phi_2)$ basis.  
If a mode has real momentum, (\ref{asymptoticsol}) ensures
that it is right moving, since the time dependence of
all modes is $\exp(iEt)$.
If a mode is non-propagating, as one mode is in the 
enhanced reflection zone,
its
momentum should be taken to be negative imaginary to give a decaying
exponential and not a growing one. Matching to the
asymptotic solutions (\ref{asymptoticsol}) 
is equivalent to imposing the boundary conditions
\begin{eqnarray}
\mbox{$\mbox{solution}_\alpha$:  } &
\vec{\phi}(x_+)  = \vec{u}_{b1}; \; &
\vec{\phi}'(x_+) = - i p_{b1} \vec{u}_{b1} \nonumber\\
\mbox{$\mbox{solution}_\beta$:  } &
\vec{\phi}(x_+)  = \vec{u}_{b2}; \; &
\vec{\phi}'(x_+) = - i p_{b2} \vec{u}_{b2} \label{solb}
\end{eqnarray}
at $x=x_+$.  Solutions to the Klein-Gordon equation (\ref{kg}) with
these complete one-point boundary conditions are easily found.  
The solutions that satisfy the symmetric phase boundary
conditions are linear combinations of solution$_\alpha$
and solution$_\beta$.  Note that if $w$ is too large, we will
run into difficulty in the enhanced
reflection zone.  We impose boundary conditions at $x_+$,
and find solution$_\alpha$ and solution$_\beta$ by evolving
toward smaller $x$.  These solutions include one mode that grows
exponentially as $x$ is reduced, until $x$ reaches $x_0$.  Therefore,
if $x_+-x_0$ is too large, the task of finding the solutions
that satisfy the symmetric phase boundary conditions involves
small differences between exponentially large quantities. We have
found that going beyond $w=40/T$ requires about 30-digit 
working precision,
and is therefore prohibitive. 

In order to obtain the desired linear combinations
of solution$_{\alpha}$ and solution$_\beta$, it is necessary
to know 
the amplitudes of the incident and reflected
components of both $\phi_1$ and $\phi_2$ 
in the symmetric phase for both solution$_\alpha$ and 
solution$_\beta$.  
At the point
$x_-$ in the symmetric phase, we denote the eigenvectors
of the mass-squared matrix defining the modes $\phi_1$ and $\phi_2$ 
by $\vec{u}_{s1}$
and $\vec{u}_{s2}$ (orthogonal because $M^2$ is Hermitian), the
masses (square roots of the corresponding
eigenvalues) by $\mu_{s1}$ and $\mu_{s2}$, and the corresponding momenta
by $p_{s1}=\sqrt{E^2-\mu_{s1}^2}$ and $p_{s2}=\sqrt{E^2-\mu_{s2}^2}$.
Then the solutions at $x_-$ will be of the form
\begin{eqnarray}
{\rm solution}_\alpha : 
\vec{\phi}(x) &=& A_{\alpha 1} e^{-i p_{s1} x} \vec{u}_{s1} +
                B_{\alpha 1} e^{+i p_{s1} x} \vec{u}_{s1} +
                A_{\alpha 2} e^{-i p_{s2} x} \vec{u}_{s2} +
		B_{\alpha 2} e^{+i p_{s2} x} \vec{u}_{s2} \nonumber\\
{\rm solution}_\beta :
\vec{\phi}(x) &=& A_{\beta 1} e^{-i p_{s1} x} \vec{u}_{s1} +
                B_{\beta 1} e^{+i p_{s1} x} \vec{u}_{s1} +
                A_{\beta 2} e^{-i p_{s2} x} \vec{u}_{s2} +
		B_{\beta 2} e^{+i p_{s2} x} \vec{u}_{s2} 
\label{a1a2b1b2}
\end{eqnarray}
where the $A$'s and $B$'s vary only slowly with $x$ 
near $x_-$ provided the mass matrix  does not change much 
on length scales comparable
to the wavelengths of the modes there. This condition
is identically satisfied for $x_-\le-w/2$, and is very well satisfied 
at larger values of $x_-$ for a wall width $25/T$.
We now write the amplitudes in (\ref{a1a2b1b2})  
in terms of $\vec\phi$ and its first derivative at $x=x_-$.
The amplitude $A_{\alpha j}$ of the incident
component of mode $j$ in solution$_\alpha$ is given by
\begin{equation}
A_{\alpha j} = e^{+i p_{sj} x_-} \vec{u}_{sj}^* \cdot
      \frac{(i p_{sj} - \partial_x)}{2 i p_{sj}} \vec{\phi}(x_-).
\end{equation}
The amplitude $B_{\alpha j}$ of the reflected component of mode $j$ 
in solution$_\alpha$ is
given by
\begin{equation}
B_{\alpha j} = e^{-i p_{sj} x_-} \vec{u}_{sj}^* \cdot
      \frac{(i p_{sj} + \partial_x)}{2 i p_{sj}} \vec{\phi}(x_-).
\end{equation}
The expressions for solution$_\beta$ are analogous.

The solutions that have incident modes in the
symmetric phase that are either purely $\phi_1$ or purely
$\phi_2$ 
can now be constructed from the amplitudes $A_{\alpha j}$,
$A_{\beta j}$, $B_{\alpha j}$ and $B_{\beta j}$.
The solution with incident $\phi_1$ and no incident $\phi_2$ is
the linear combination of solution$_\alpha$ and solution$_\beta$
for which the $A_2$ terms cancel, namely
\begin{equation}
\mbox{$\mbox{solution}_1$:  }
     A_{\beta  2} \mbox{ solution}_\alpha -
     A_{\alpha 2} \mbox{ solution}_\beta \ .
\end{equation}
Solution$_1$ can be written in the form (\ref{a1a2b1b2}),
with coefficients
\begin{eqnarray}
A_{11} &=& A_{\beta 2}  A_{\alpha 1} -
	   A_{\alpha 2} A_{\beta 1} \nonumber \\
A_{12} &=& A_{\beta 2}  A_{\alpha 2} -
	   A_{\alpha 2} A_{\beta 2} = 0 \nonumber \\
B_{11} &=& A_{\beta 2}  B_{\alpha 1} -
	   A_{\alpha 2} B_{\beta 1} \nonumber \\
B_{12} &=& A_{\beta 2}  B_{\alpha 2} -
	   A_{\alpha 2} B_{\beta 2}\ .
\end{eqnarray}
These coefficients are the amplitudes of the incident
and reflected modes in solution$_1$.
Similarly, the solution with incident $\phi_2$ and no incident
$\phi_1$ is
\begin{equation}
\mbox{$\mbox{solution}_2$:  }
     A_{\beta  1} \mbox{ solution}_\alpha -
     A_{\alpha 1} \mbox{ solution}_\beta
\end{equation}             
with amplitudes
\begin{eqnarray}
A_{21} &=& A_{\beta 1}  A_{\alpha 1} -
	   A_{\alpha 1} A_{\beta 1} = 0 \nonumber \\
A_{22} &=& A_{\beta 1}  A_{\alpha 2} -
           A_{\alpha 1} A_{\beta 2} \nonumber \\
B_{21} &=& A_{\beta 1}  B_{\alpha 1} -
	   A_{\alpha 1} B_{\beta 1} \nonumber \\
B_{22} &=& A_{\beta 1}  B_{\alpha 2} -
           A_{\alpha 1} B_{\beta 2}.
\end{eqnarray}
We now have all the ingredients necessary to construct the
reflection coefficients $R_{ij}$.  (Note that as
shown in Section II, $R_{\overline{ij}}=R_{ji}$.)

The reflection coefficient $R_{ij}$  
is defined the be the ratio of the
reflected $\phi_j$ current into the symmetric phase
to the incident $\phi_i$ current from the symmetric phase.  The
current represented by a solution $\phi(x)$ is
$i(\partial_x \phi^*) \phi - i\phi^* (\partial_x \phi)$. For a
solution $A e^{i(E t-p x)}$, this current is $2 |A|^2 \mbox{Re }(p)$.
Hence the reflection coefficients are:
\begin{eqnarray}
R_{11} &=& \left|\frac{B_{11}}{A_{11}}\right|^2
           \frac{\mbox{Re }p_{s1}}{\mbox{Re }p_{s1}}
        =  \left|\frac{A_{\beta 2} B_{\alpha 1} - A_{\alpha 2} B_{\beta 1}}
                      {A_{\beta 2} A_{\alpha 1} - A_{\alpha 2} A_{\beta 1}}
           \right|^2 \nonumber \\
R_{12} &=& \left|\frac{B_{12}}{A_{11}}\right|^2
	   \frac{\mbox{Re }p_{s2}}{\mbox{Re }p_{s1}}
        =  \left|\frac{A_{\beta 2} B_{\alpha 2} - A_{\alpha 2} B_{\beta 2}}
                      {A_{\beta 2} A_{\alpha 1} - A_{\alpha 2} A_{\beta 1}}
           \right|^2 \frac{\mbox{Re }p_{s2}}{\mbox{Re }p_{s1}} \nonumber \\
R_{21} &=& \left|\frac{B_{21}}{A_{22}}\right|^2
	   \frac{\mbox{Re }p_{s1}}{\mbox{Re }p_{s2}}
	=  \left|\frac{A_{\beta 1} B_{\alpha 1} - A_{\alpha 1} B_{\beta 1}}
		      {A_{\beta 1} A_{\alpha 2} - A_{\alpha 1} A_{\beta 2}}
           \right|^2 \frac{\mbox{Re }p_{s1}}{\mbox{Re }p_{s2}} \nonumber \\
R_{22} &=& \left|\frac{B_{22}}{A_{22}}\right|^2
	   \frac{\mbox{Re }p_{s2}}{\mbox{Re }p_{s2}}
	=  \left|\frac{A_{\beta 1} B_{\alpha 2} - A_{\alpha 1} B_{\beta 2}}
		      {A_{\beta 1} A_{\alpha 2} - A_{\alpha 1} A_{\beta 2}}
           \right|^2. \label{r}
\end{eqnarray}

\end{document}